\documentclass[screen=true, sigconf]{acmart}

\setcopyright{acmlicensed}
\acmPrice{15.00}
\acmDOI{10.1145/3368089.3409711}
\acmYear{2020}
\copyrightyear{2020}
\acmSubmissionID{fse20main-p316-p}
\acmISBN{978-1-4503-7043-1/20/11}
\acmConference[ESEC/FSE '20]{Proceedings of the 28th ACM Joint European Software Engineering Conference and Symposium on the Foundations of Software Engineering}{November 8--13, 2020}{Virtual Event, USA}
\acmBooktitle{Proceedings of the 28th ACM Joint European Software Engineering Conference and Symposium on the Foundations of Software Engineering (ESEC/FSE '20), November 8--13, 2020, Virtual Event, USA}

\usepackage[utf8]{inputenc} 
\usepackage{amsmath,amssymb,amsfonts}
\usepackage{arydshln}
\usepackage[inline]{enumitem}
\usepackage{xspace}
\usepackage{booktabs}
\usepackage{listings}
\usepackage{url}
\usepackage{enumitem}
\usepackage{flushend}

\setlist[description]{labelindent=0pt,style=multiline,leftmargin=1cm}

\begin{document}

\title{Selecting Third-Party Libraries: The Practitioners' Perspective}

\author{Enrique~Larios Vargas}
	\affiliation{
	\institution{Software Improvement Group}
	\country{The Netherlands}
	}
	\email{e.lariosvargas@sig.eu}
	
	\author{Maurício~Aniche}
	\affiliation{\institution{Delft University of Technology}
	\country{The Netherlands}
	}
	\email{M.FinavaroAniche@tudelft.nl}
	
	\author{Christoph~Treude}
	\affiliation{\institution{University of Adelaide}
	\country{Australia}
	}
	\email{christoph.treude@adelaide.edu.au}
	
	\author{Magiel~Bruntink}
	\affiliation{\institution{Software Improvement Group}
	\country{The Netherlands}
	}
	\email{m.bruntink@sig.eu}
	
	\author{Georgios~Gousios}
	\affiliation{\institution{Delft University of Technology}
	\country{The Netherlands}
	}
	\email{G.Gousios@tudelft.nl}	

\newcommand{\factor}[1]{\textit{#1}}
\newcommand{\source}[2]{\textit{#1~\textsuperscript{(\em #2)}}}
\newcommand{\block}[2]{\textit{#1~\textsuperscript{(#2)}}}

\newcommand{\fix}[1]{\textcolor{red}{#1}}
\newcommand{\added}[1]{\textcolor{blue}{#1}}

\newcommand{\cooltitle}[1]{\vspace{1mm}\noindent\textbf{#1}}

\newcommand{\numberoffactors}{26\xspace}
\newcommand{\numberofsurveys}{115\xspace}
\newcommand{\numberofinterviews}{16\xspace}
\newcommand{\numberofnewfactors}{8\xspace}

\newcommand{\hist}[1]{(\includegraphics[height=2.5mm]{#1})}
\newcommand{\histt}[1]{\includegraphics[height=2.5mm]{#1}}

\begin{abstract}
The selection of third-party libraries is an essential element of virtually any software development project. However, deciding which libraries to choose is a challenging practical problem. Selecting the wrong library can severely impact a software project in terms of cost, time, and development effort, with the severity of the impact depending on the role of the library in the software architecture, among others. Despite the importance of following a careful library selection process, in practice, the selection of third-party libraries is still conducted in an ad-hoc manner, where dozens of factors play an influential role in the decision.

In this paper, we study the factors that influence the selection process of libraries, as perceived by industry developers. To that aim, we perform a cross-sectional interview study with \numberofinterviews developers from 11 different businesses and survey \numberofsurveys developers that are involved in the selection of libraries. We systematically devised a comprehensive set of \numberoffactors technical, human, and economic factors that developers take into consideration when selecting a software library. Eight of these factors are new to the literature. We explain each of these factors and how they play a role in the decision. Finally, we discuss the implications of our work to library maintainers, potential library users, package manager developers, and empirical software engineering researchers.

\end{abstract}



\maketitle

\section{Introduction}
\label{sec-intro}

The use of third-party libraries\footnote{We use the term ``library'' to denote frameworks, libraries, and APIs.} is a recurrent practice among practitioners to speed up the development of software systems and, as a consequence, to reduce its costs~\cite{mojica2014}. Software libraries provide developers with customized functionality which is useful while developing software~\cite{li2017}. Developers do not need to ``reinvent the wheel'', but rather seek libraries that suit their purpose~\cite{nguyen2020}. 

The selection of the right library to adopt is a demanding task for developers. The number of factors that should be taken into account during the selection process is significant. For example, the adoption of a library that goes against the existing architectural decisions of the software system might increase its adoption time; a library that is easy to use but not performant enough might have an impact on the quality of the system; a library that has good documentation and active community, but with which the team has had a bad experience, might not be the best possible choice. On top of that, the number of libraries available is overgrowing~\cite{xu2020}. While such growth gives developers more options, it also complicates their selection process.

The process that developers follow to select a software library is often ad-hoc, and we have no clear evidence that these decisions are quality-driven~\cite{zaimi2015}. Indeed, different software practitioners might consider different criteria when choosing a library, e.g., how much they trust it~\cite{kula2015}, and how popular the library is in its ecosystem~\cite{yano2015,mileva2009,mileva2010,lima2019}. However, we argue that \textit{the lack of a systematic approach may lead software developers to choose libraries arbitrarily, without considering the consequences of their decisions}. 

Being aware of the criteria applied by practitioners and how these factors affect the selection of third-party libraries would enable potential new users to make systematic decisions, side-stepping biases that could lead to suboptimal choices~\cite{arnott2005,milkman2009}. It would also enable researchers to improve the state-of-the-art in software engineering tools. It is vital to have a clear understanding of the actual criteria and decision process before building tools to assist practitioners. Our work builds the foundation for an in-depth understanding of how developers select their libraries. We conjecture that together with appropriate tools, supported by large-scale data, high transparency, and near-real-time feedback~\cite{larios2018}, practitioners will be capable of making data-driven (and thus, more assertive) decisions about which libraries to adopt.  

The software engineering research community has been investigating the selection process of libraries and, more specifically, the factors that developers take into consideration (e.g., ~\cite{delamora2018a, delamora2018b, pano2016, xu2020, hora2015, abdalkareem2017}; see Table~\ref{tab:factors-related-work} for a full list of factors found in related work). However, the spectrum of factors influencing the decision is, as we see in the table mentioned above, highly diverse and dependent on the context, leaving the door open for more research in this area.

In this paper, we aim to understand what are the factors that influence the library selection process, as perceived by industry practitioners. We present the results of a cross-sectional interview study and a survey. We first interviewed \numberofinterviews software practitioners from 11 companies in different domains, such as banking, consulting, and oil and gas. We used semi-structured interviews to collect data about the criteria that software practitioners follow when selecting libraries. Subsequently, we challenged the interview findings with an additional \numberofsurveys participants from the industry and open source communities. 

Our study leads to the following contributions: 

\begin{enumerate}[label=(\alph*)]

\item A comprehensive set of \numberoffactors factors (\numberofnewfactors of them being new to the literature) that influence the selection process of software libraries, after interviewing \numberofinterviews developers from different businesses, and validating them with \numberofsurveys survey respondents.
    
\item A discussion of the implications of our work on library maintainers (how they can increase the adoption rate of their libraries), potential library users (how they can make their selection process more systematic), package manager developers (how these tools can support developers in selecting libraries), and researchers (suggestions for future work).
\end{enumerate}

\section{Research Method}
\label{sec-method}

The goal of this study is to \textbf{understand the factors that practitioners take into account when selecting software libraries}. To that aim, our research methodology consisted of a cross-sectional interview study~\cite{creswell2013research} and a survey. Our study design included the following three steps:
\begin{enumerate*}[label=(\roman*)]
    \item conduct semi-structured interviews with software practitioners involved in the selection of software libraries,
    \item build and iteratively refine the factors book through several discussions among the researchers and member checking sessions, and
    \item conduct a survey to challenge and validate the findings.
\end{enumerate*}

In the following sections, we explain our methodology in detail. Private information from interviewees and companies has been anonymized. The authors do not have the participants' authorization to make the raw interview scripts available as they contain private information.

\subsection{Interviews}

Our first step was to conduct semi-structured interviews. Interview-based research is appropriate for gathering in-depth descriptions of experiences, observations, and assessments~\cite{hiller2004}. Semi-structured interviews, in contrast to entirely structured interviews, tend to encourage participants to freely share their thoughts and enable researchers to explore new ideas based on the answers of the interviewee~\cite{hove2005experiences}. 

The goal of the interviews was to collect the developers' stories, experiences, and challenges on how they decided to select (or not select) a software library.  Our interview guide focused on the following topics:
\begin{enumerate}[label=(\alph*)]
    \item the practitioner's (i.e., the interviewee's) perception of influential factors for library selection, 
    \item how these factors play an influential role in the selection process, 
    \item the sources of information (e.g., websites, work colleagues) that are commonly used by practitioners while selecting a library, and
    \item the challenges that practitioners commonly face in the selection process.
\end{enumerate}

To foster the discussion and make it always about a recent experience the participants had, we often reminded them to talk about their most recent experiences in selecting third-party libraries. The influential factors for library selection emerged out of the stories that practitioners were telling. We avoided their personal opinions when disconnected from a real case. 

We recruited participants who have been part of software library selection processes in the past. The initial pool of interviewees came from convenience sampling: the authors of this paper invited their industry contacts to join the study. Other participants were then collected by snowballing (P10, P13, P14, P15, P16). We had strict selection criteria, where participants should meet the following conditions: 
\begin{enumerate*}[label=(\alph*)]
	\item they should have at least four years of experience, and
	\item they had to be intensively involved in the selection of software libraries.
\end{enumerate*}

\begin{table*}[]
\small
\centering
\begin{tabular}{lllrlr}

\toprule
\textbf{Interviewee} & \textbf{Company} & \textbf{Business} & \textbf{Company size}                              & \textbf{Role/Function}                                 & \textbf{Years of Exp.} \\

\midrule

P1    & C        & Banking  & 10000+                            & Front-End Software Developer                  & 4             \\
P2    & B        & Education & 5000+                & Software Developer                            & 8             \\
P3    & -        & Research & self-employed                & Scientific developer                                   & 4            \\
P4    & A        & Consultancy & 100+                           & Senior Technical Consultant                   & 12            \\
P5    & A        & Consultancy & 100+                          & Head of Software Development Team             & 15            \\
P6    & B        & E-Learning & 50+               & Software Developer       & 8             \\
P7   & A        & Consultancy & 100+                          & Senior Technical Consultant                   & 12            \\
P8   & D        & CRM and Digital Process Automation & 5000+    & Software Developer                            & 4             \\
P9   & E        & Full stream oil and gas solutions & 10000+                       & Software Developer                            & 7             \\
P10   & F        & Busines to Business Framework & 100+         & Senior Software Engineer                      & 8             \\
P11   & G        & Automated visual monitoring solutions & 50+ & Senior Software Engineer                      & 13            \\
P12   & H        & Microscopic Imaging & 50+                  & Senior Scientific Software Engineer           & 15            \\
P13   & I        & All-in-one mobile travel platform & 50+     & Senior Software Engineer                      & 14            \\
P14   & I        & All-in-one mobile travel platform & 50+     & Senior Software Engineer/Solution Architect   & 10            \\
P15   & J        & Embedded Systems & 1000+                      & Senior Software Engineer                      & 13            \\
P16   & K        & Consultancy & 10000+                          & Software Engineer/Technical Architect         & 8             \\

\bottomrule

\end{tabular}
\vspace{0.02in}
\caption{Profile of our participants (N=\numberofinterviews). Companies are anonymized.}
\label{tab:participants}
\vspace{-5mm}

\end{table*}

We conducted interviews until the authors came to an agreement that theoretical saturation was reached. According to Strauss and Corbin~\cite{strauss1997grounded}, sampling should be discontinued once the already collected data is considered sufficiently dense, and data collection no longer generates new information. 

In the end, we interviewed \numberofinterviews software practitioners (the participants are identified as P1 -- P16 throughout this paper) from 11 different companies that work in 9 different fields. Each interview lasted around 45 minutes, producing a total of around 15 hours of recorded audio. The experience of participants in software development ranged from 4 to 15 years (median=10 years). In Table~\ref{tab:participants}, we detail our participants.

We iteratively built a factors book containing the influential factors mentioned by the interviewees. 
After each interview, one of the researchers (the first author of this paper) transcribed the audio recording and performed \textit{open coding} of the interview data, using the MaxQDA software. The open coding process consisted of understanding the factors that the interviewee takes into account when selecting a software library and attaching a code to it. Then, two researchers (the first two authors of the paper) analyzed the factors that emerged from that participant, compared them with the findings from previous participants, and refined the factors book.

Finally, we performed member-checking sessions to validate the factors book with six of the interviewees. Ten participants were randomly selected and invited for the member-checking stage. Six of them were available to participate. The goal of member-checking was to enable participants to analyze the factors book critically and to provide feedback. The sessions were conducted in no specific order but according to the availability of our participants. We repeated this process until we considered the factors book mature enough. To reach this decision, we also took into account the amount of feedback received in each member-checking session (similar to saturation, we stopped when the participants' feedback was not producing any new insights). Throughout this process, we devised ten different versions of the factors book. The first four versions resulted from four rounds of discussions among the researchers, and six other versions were generated after each member-checking session. The final version of the factors book resulted in \textbf{\numberoffactors influential factors} grouped in technical, human, and economic aspects of the library.    

\subsection{Survey}

We designed a survey to validate the findings obtained from the interviews. Each question was related to a factor in our factors book. Participants had to indicate their perception on whether each factor had low or high influence on the selection process of a library. Answer options ranged from no influence to high influence, in a 4-points Likert scale (i.e., no influence, low influence, moderate influence, high influence). We also allowed respondents to answer ``I do not know'' or ``does not apply''. 

To avoid possible confusion from participants, for each of the factors, we do not show the ``raw factor'', but a sentence that explains it. We refined the sentences by means of pilots with four participants (these data points were later discarded and are not part of our results). To complement our list of factors, we also asked survey participants, through an open-ended question, what other factors have influenced their selection of libraries, as a strategy to ensure the quality of the interview data. SurveyGizmo's estimated time to answer the survey was seven minutes. 

Aiming to recruit industry professionals, we shared the survey on four social networks: Twitter, LinkedIn, Reddit, and Facebook. We informed participants that we would reward their answers by donating a total of 150 Euros, proportionally divided to the three different institutions that participants could choose (Médecins Sans Frontières (MSF), UNICEF, and the Red Cross).

We ran the survey in two different languages (English and Brazilian Portuguese), obtaining 177 responses in total. However, some participants did not answer all questions which were discarded from the analysis, leaving a total of \numberofsurveys complete responses. Our survey respondents were from 4 different continents (42 from Europe, 70 from North and South America, 1 from Asia, and 2 from Oceania). Respondents vary regarding how many years of work experience they have, with a range between 1 and 30 years of work experience (median = 9 years). Concerning the type of business where our respondents work, 34\% indicated to be working in web development, 13\% in consulting, 12\% in education, 11\% in banking software development, 6\% in scientific development, 3\% in mobile development, 2\% in open source development, and 19\% in others. Finally, concerning respondents' role in their software projects, 36\% indicated to be technical lead or software architects, 42\% software developers, 11\% researchers, 5\% project or product managers, and 6\% other roles. 

The survey form, the mapping between the survey questions and the factors in the factors book, as well as the answers we received, are available in our online appendix~\cite{appendix}. 
\section{Results}
\label{sec-results}

\newcommand{\here}{$\bullet$}

\begin{table*}[]
\footnotesize
\resizebox{\textwidth}{!}{
\begin{tabular}{lccccccccccccccccrrrrcc}
\toprule

& \textbf{P1} & \textbf{P2} & \textbf{P3} & \textbf{P4} & \textbf{P5} & \textbf{P6} & \textbf{P7} & \textbf{P8} & \textbf{P9} & \textbf{P10} & \textbf{P11} & \textbf{P12} & \textbf{P13} & \textbf{P14} & \textbf{P15} & \textbf{P16} &  & \textbf{H} & \textbf{L+M} & \textbf{N} & \textbf{R} & \textbf{U/P} \\
\midrule

\textbf{Technical factors} & & & & & & & & & & & & & & & & & & & & \\ \midrule

\hspace{1mm}\textbf{Software system} & & & & & & & & & & & & & & & & & & & & \\

\hspace{2mm}Brown- or green-field & \here & \here & \here & \here & \here & \here & & & & & \here & & \here & & & \here & \histt{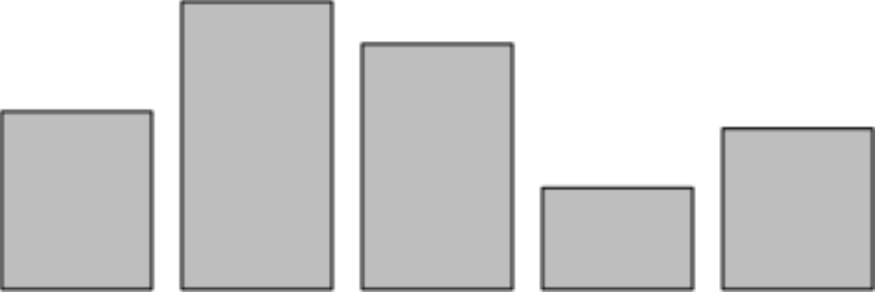} & 10\% & 55\% & 18\% & \checkmark & * \\ \hdashline
\hspace{1mm}\textbf{Functionality} & & & & & & & & & & & & & & & & & & & & \\
\hspace{2mm}Size and complexity & \here & \here & & \here & \here & & \here & & & & \here & & & & & & \histt{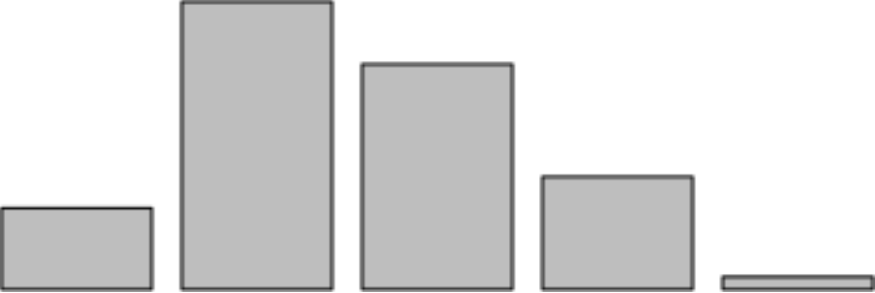} & 16\% & 71\% & 11\% & & P \\
\hspace{2mm}Fit for purpose & \here & \here & & \here & \here & & \here & \here & & \here & & & \here & & \here & \here & \histt{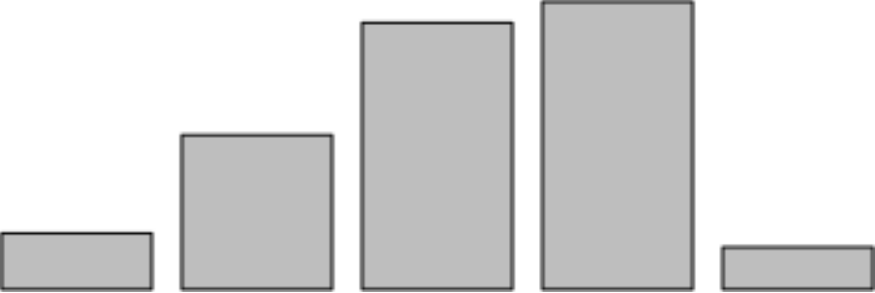} & 36\% & 52\% & 7\% & & U \\ \hdashline
\hspace{1mm}\textbf{Quality} & & & & & & & & & & & & & & & & & & & \\
\hspace{2mm}Alignment w/ architecture & & & & & \here & & \here & & \here & \here & & & \here & & & & \histt{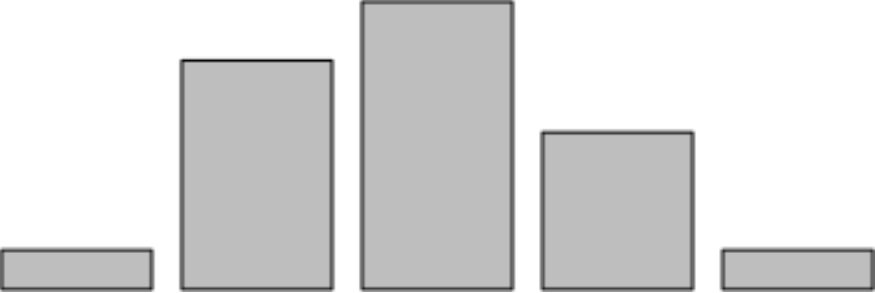}  & 21\% & 69\% & 5\% & & U\\
\hspace{2mm}Usability & & \here & \here & & & & \here & \here & \here & & \here & & & & & \here & \histt{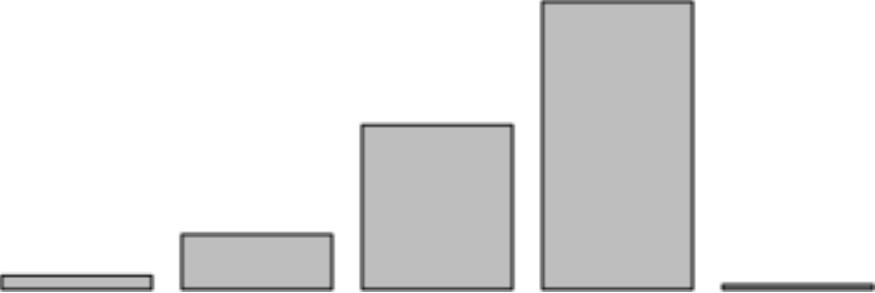}  & 55\% & 42\% & 3\% & & P\\
\hspace{2mm}Documentation & \here & \here & \here & & \here & \here & \here & \here & \here & \here & & & & & \here & \here & \histt{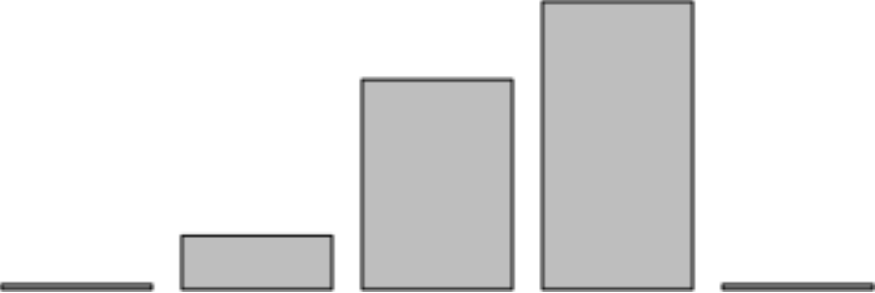} & 51\%  & 47\% & 1\% & & U/P \\
\hspace{2mm}Security  & \here & \here & & & \here & & \here & & & & \here & & \here & & \here & \here & \histt{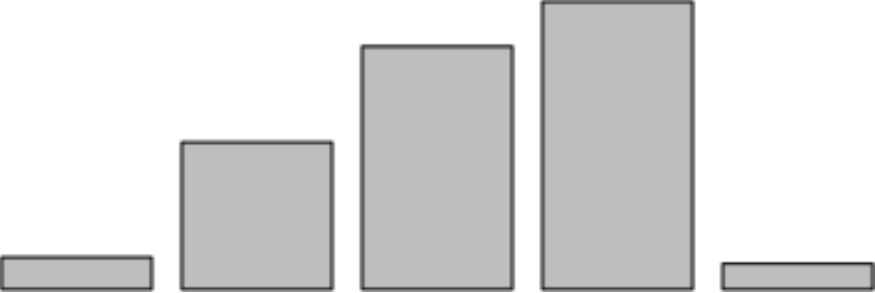} & 39\% & 53\% & 4\%  & & U/P \\
\hspace{2mm}Performance & & \here & & & & & & \here & \here & & & \here & & & \here & & \histt{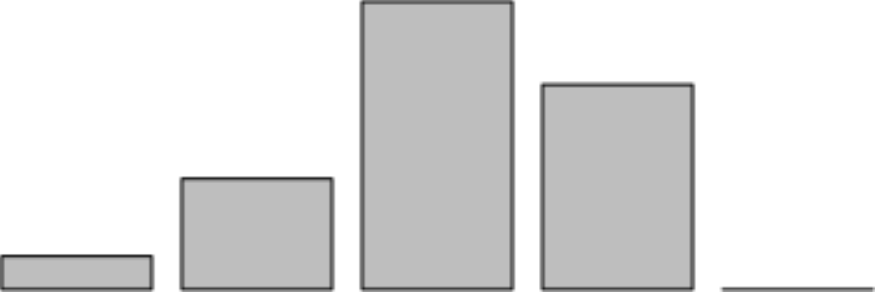} & 32\% & 63\% & 5\% & & P \\
\hspace{2mm}Well tested & \here & & & & & & & & & \here & \here & & & & \here & & \histt{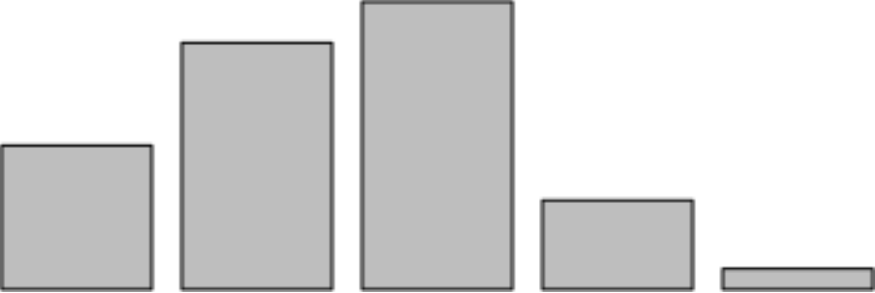} & 11\% & 68\% & 18\% & & U \\ \hdashline
\hspace{1mm}\textbf{Release} & & & & & & & & & & & & & & & & & \\
\hspace{2mm}Active maintenance & \here & \here & \here & & \here & & \here & & \here & \here & \here & \here & \here & & & \here & \histt{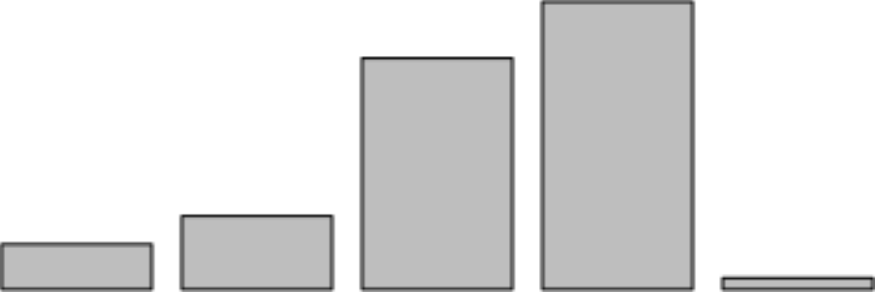} & 44\% & 47\% & 7\% & & U \\
\hspace{2mm}Maturity and stability & \here & \here & & & & \here & & & \here & \here & & & & & \here & & \histt{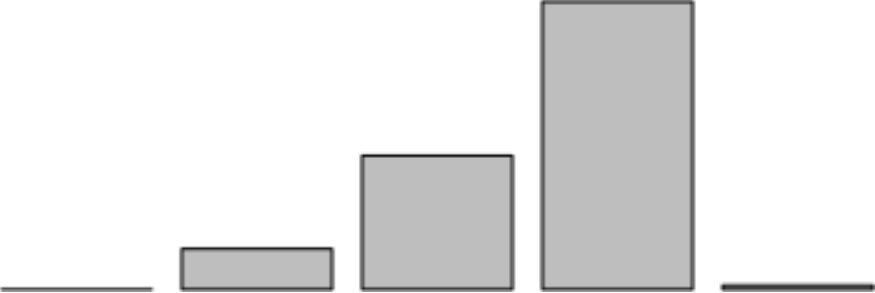}  & 62\% & 37\% & 0\% & & U\\
\hspace{2mm}Release cycle frequency & \here & & & \here & & & \here & & & & \here & & \here & & & & \histt{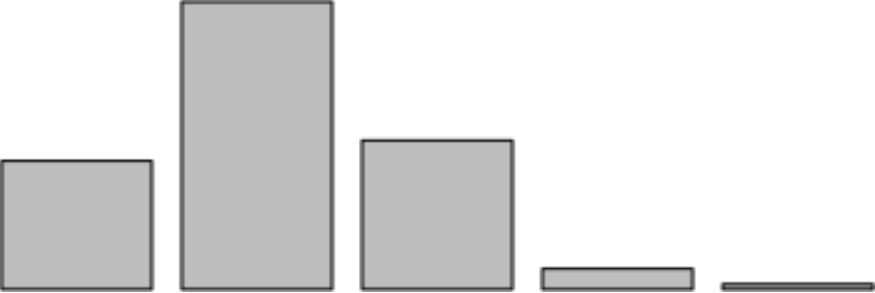} & 3\% & 74\% & 22\% & & U \\

\midrule

\textbf{Human factors} & & & & & & & & & & & & & & & & & \\ \midrule

\hspace{1mm}\textbf{Stakeholders} & & & & & & & & & & & & & & & & & \\
\hspace{2mm}Customers & & \here & & & & & & & & & & & & \here & & \here & \histt{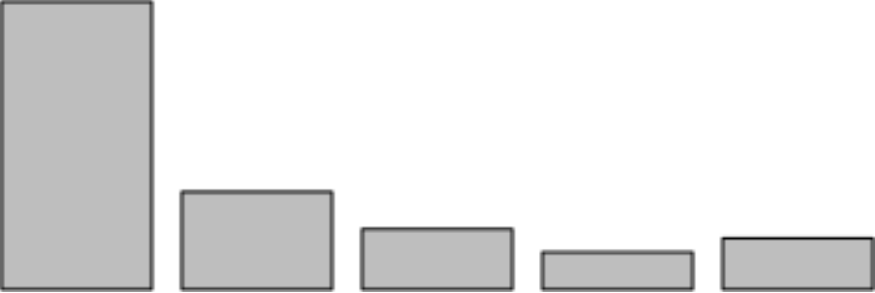} & 7\% & 30\% & 54\% & & U/P\\
\hspace{2mm}Other teams & \here & \here & & & & & & \here & & \here & & & & & \here & & \histt{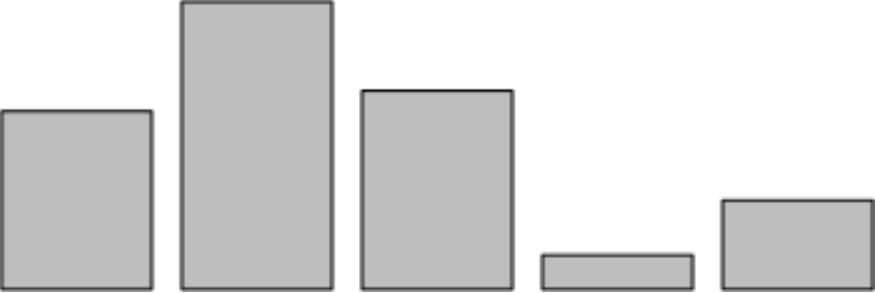} & 4\% & 62\% & 23\% & \checkmark & U/P\\
\hspace{2mm}Project/product managers & & & & & & & & & & \here & & & & & & \here & \histt{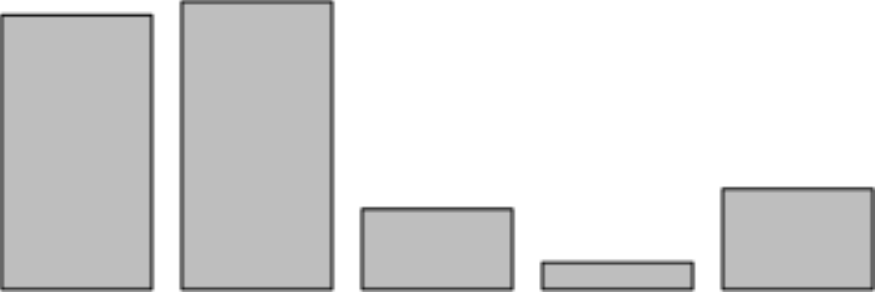} & 3\% & 48\% & 36\% & \checkmark & U/P\\
\hspace{2mm}Development team & \here & & & \here & \here & \here & \here & \here & \here & \here & \here & \here & \here & \here & \here & \here & \histt{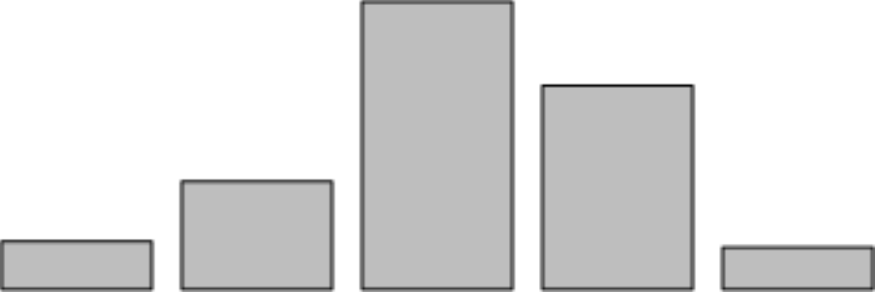} & 30\% & 57\% & 7\% & & U/P \\ \hdashline

\hspace{1mm}\textbf{Organization} & & & & & & & & & & & & & & & & & \\
\hspace{2mm}Type of industry & & & & & & & & & & & \here & & & \here & & \here & \histt{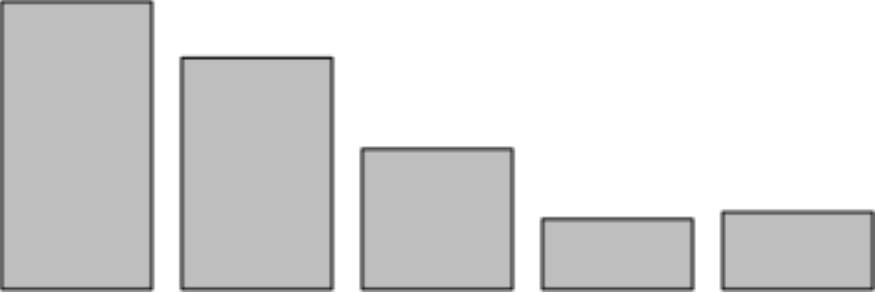} & 9\% & 46\% & 36\% & \checkmark & U/P\\
\hspace{2mm}Culture and policies & \here & & & & & & & & & \here & & & & \here & \here & \here & \histt{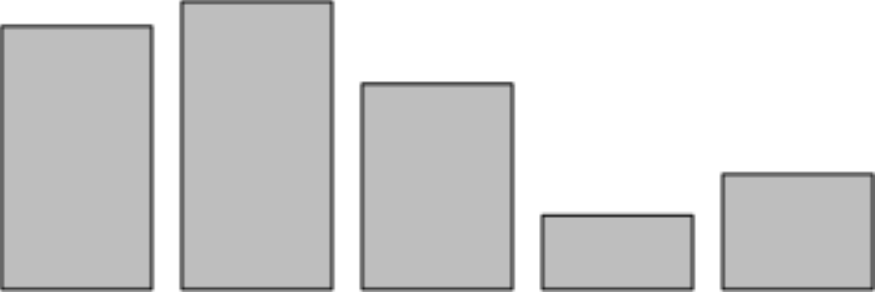} & 8\% & 52\% & 28\% & \checkmark & U/P\\
\hspace{2mm}Management and strategy & & & & & & & & & & \here & \here & & & & & & \histt{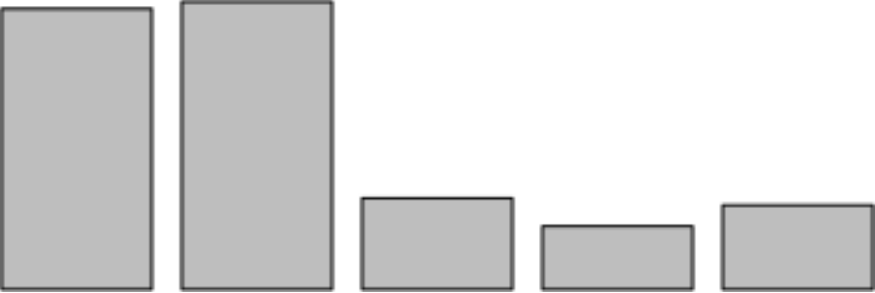} & 8\% & 47\% & 35\% & \checkmark & U/P \\ \hdashline
\hspace{1mm}\textbf{Individual} & & & & & & & & & & & & & & & & & \\
\hspace{2mm}Self-perception & & & \here & & & \here & & & & \here & & & & & & & \histt{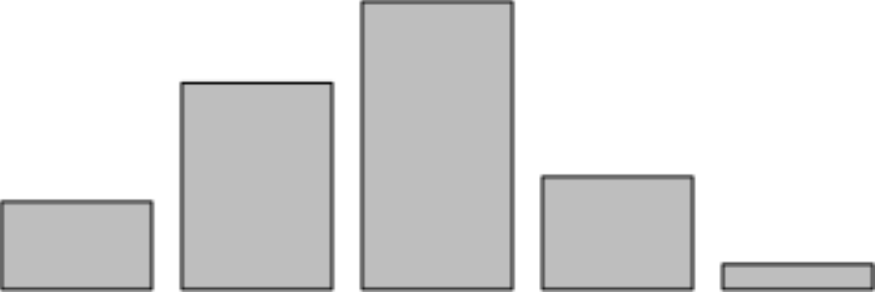}  & 16\% & 69\% & 12\% & \checkmark & U/P\\
\hdashline
\hspace{1mm}\textbf{Community}  & & & & & & & & & & & & & & & & & \\
\hspace{2mm}Experience & \here & \here & & \here & & \here & & & & \here & \here & & \here & \here & & \here & \histt{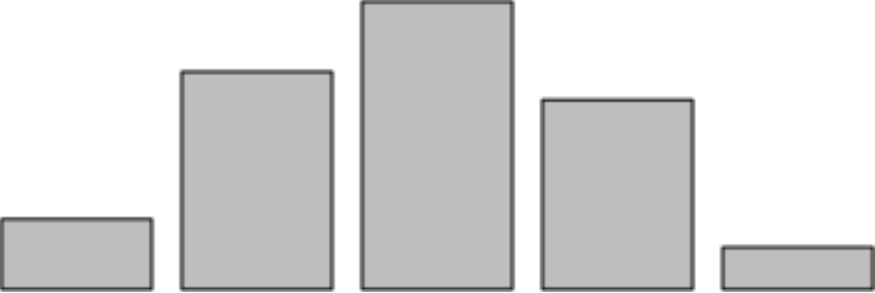} & 23\% & 63\% & 9\%  & & U/P \\
\hspace{2mm}Activeness & & & \here & & & & \here & \here & \here & \here & & & & \here & \here & & \histt{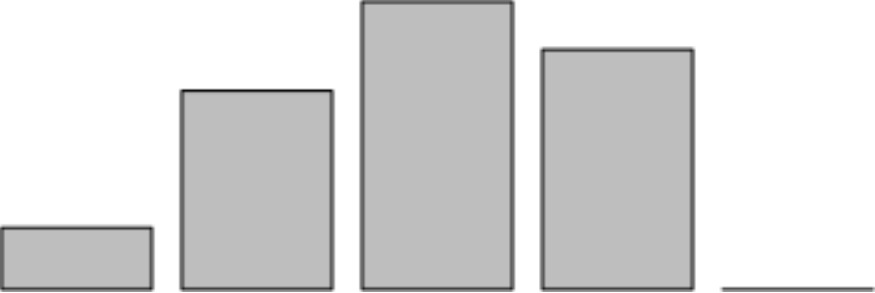}  & 30\% & 62\% & 8\% & & U/P \\
\hspace{2mm}Popularity & \here & \here & \here & \here & \here & \here & \here & & \here & \here & \here & \here & \here & \here & \here & \here & \histt{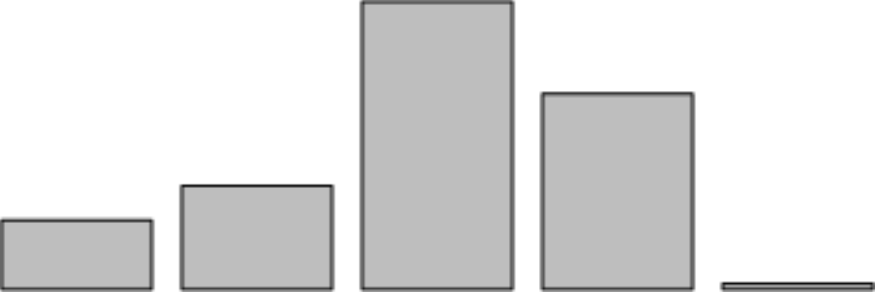} & 30\% & 59\% & 10\% & & U/P \\

\midrule

\textbf{Economical factors}  & & & & & & & & & & & & & & & & &  \\ \midrule
\hspace{1mm}\textbf{Total cost of ownership} & & & & & & & & & & & & & & & & & & \\
\hspace{2mm}Time and budget & & & & & \here & & & & \here & \here & & & & \here & \here & \here & \histt{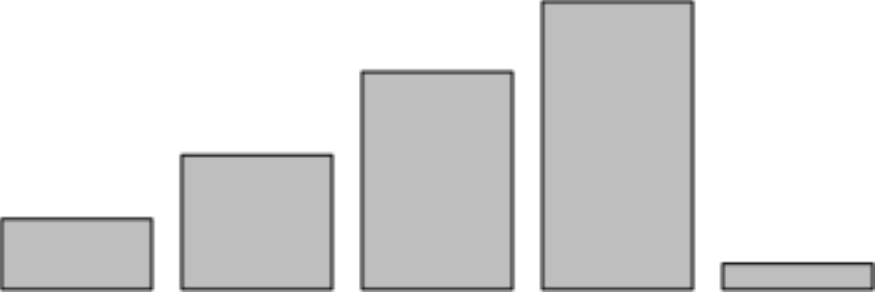} & 39\%  & 48\% & 10\% & & U/P \\
\hspace{2mm}License & \here & \here & & \here & \here & & & \here & \here & \here & & \here & \here & \here & \here & \here & \histt{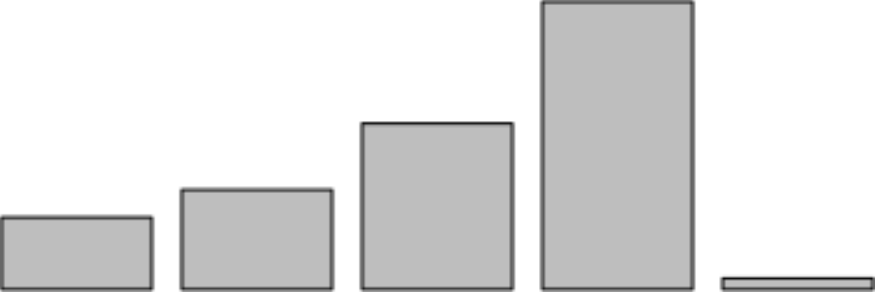} & 45\% & 42\% & 11\% & & U \\ \hdashline
\hspace{1mm}\textbf{Risk} & & & & & & & & & & & & & & & & & & \\
\hspace{2mm}Risk assessment & & \here & & \here & & & & & & \here & \here & & & \here & & & \histt{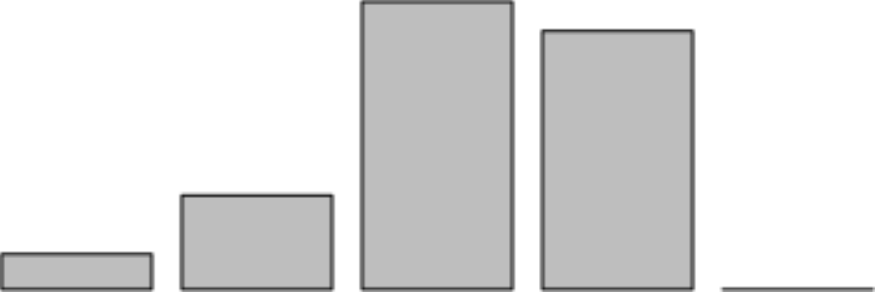} & 38\% & 57\%  & 5\% & \checkmark & P\\

\bottomrule
\end{tabular}
}
\caption{The \numberoffactors factors, per interviewee (P1-P16). 
Bar plots represent the responses of our \numberofsurveys survey participants (no influence, low influence, moderate influence, high influence, and does not apply, respectively).
Columns ``H'' (high influence), ``L+M'' (low or moderate influence), and ``N'' (no influence) indicate their percentage. The percentage of respondents who choose ``NA'' (Not Applicable) is not shown in the table. R indicates whether the factor is a new contribution to the existing literature. U/P indicates whether the factor is evaluated during the U (up-front) or P (prototyping) phase.}
\label{tab:all-factors}
\end{table*}

We group the aspects that practitioners take into account when selecting libraries into three categories of factors: 
 \begin{enumerate*}[label=(\alph*)]
 	\item technical (\textit{functionality}, \textit{quality}, \textit{type of project}, and \textit{release process}),
 	\item human (\textit{stakeholders}, \textit{organization}, \textit{individual}, and \textit{community}), and 
 	\item economic factors (\textit{total cost of ownership (TCO)} and \textit{risk}). 
 \end{enumerate*}
In Table~\ref{tab:all-factors}, we show the entire landscape of factors, interviewees that mention that factor, and the percentage of survey participants that indicate certain levels of influence for a given factor when selecting a library.

In the following, we discuss each of the three categories in detail (Sections~\ref{sec:technical-factors}, \ref{sec:human-factors}, and \ref{sec:economical-factors}), as well as the role of these factors in the overall selection process (Section~\ref{sec:interplay}). 
\subsection{Technical Factors}
\label{sec:technical-factors}

Practitioners consider three technical aspects when selecting a library: \textit{the characteristics of its releases and release process}, \textit{its quality attributes}, and \textit{the functionalities it provides}. Moreover, as a characteristic of the software that will make use of the library, the project being a \textit{green-field} development project (i.e., it starts from scratch, with no restrictions from existing software or previous architectural decisions) or a \textit{brown-field} development project (i.e., it contains constraints that could have been imposed by existing software) is also a technical factor taken into account by practitioners. 
\subsubsection{\textbf{The characteristics of the releases and the release process of the library.}}
\label{subsubsec:release}

Practitioners explain that every time a new library is adopted in a project, a new dependency on someone else's code is established for the long term. Decision-makers trust in libraries that provide positive signs of being active for a long period. Therefore, characteristics that are associated with a library's release provide developers with indicators to predict for how long a library will be alive, i.e., whether the library is actively maintained, whether it offers long-term support, its maturity and steady evolution, how often it releases new versions, whether there is a recent release, and how stable it is. 
 
The \factor{active maintenance} of a library is perceived as a positive sign by practitioners (P1-P3, P5, P7, P9-P13, P16). A continuous decrease in a library's maintenance activity provides indications of a short life span of the library. For instance, P13 points out that ``\textit{If a project is not active for one year, I would think that is a dead project, and I will discard it}''. Given that there is not a single way of measuring whether a library is actively maintained, practitioners rely on different metrics, e.g., by investigating the volume of contributions in the library's repository (P1), by exploring how regularly the library is updated (P3, P5), by observing if its most recent release has happened recently (P7, P11, P16), or by seeing whether contributors are actively working on new features or fixing bugs (P9). In our survey, we note that 44\% of the participants perceive that a library being actively maintained is a highly influential factor in the selection process. Only 7\% consider it not to have any influence in the selection process~\hist{icons/hmaintenance.pdf}.

The concern for long-term support forces participants to examine other characteristics such as the \factor{maturity and stability} of the library. P6 points out that examing stability is fundamental for understanding if it is necessary to update the library:
``\textit{I care about stability in the sense if I need to update the library frequently. Because I don't want to. Hibernate, for example, you can spend one year using an old version, and that is ok}''.
We observe that participants do not have a standard way to measure the stability of a library; examining the volume of bugs identified per release (P10), and whether the library is getting enough ``momentum'' (P2, P4, and P12) were some of the techniques our participants mentioned.
62\% of our survey respondents perceive \factor{maturity and stability} as a highly influential factor, and none of the respondents believes this factor does not have any influence in the selection process~\hist{icons/hmaturitystability.pdf}. 

The release process itself, in particular, the \factor{release frequency} of a library, is also a factor that participants take into account (P1, P4, P7, P11, P13). In general, we observe that short release cycles serve as a positive factor for practitioners to select a library. However, how practitioners take this factor into account might vary according to the type of library they are about to select. For example, P1 perceives libraries with short release cycles in a positive manner. An exception is made when the library is considered to be a ``big one'', for example, a complex persistence framework such as Hibernate (which is ``bigger'' compared to a ``simple'' string utils library). In their words: ``\textit{I would trust more in a library which has shorter release cycles, but if it is a big library I would not prefer it because that means that I will need to update libraries to a recent version frequently}''.  Moreover, P13 points out that ``\textit{As long as they are releasing, one month, or two months, I think it is better than being updated like in a year. At least I can see that the project is alive}''. Interestingly, only 3\% of our survey respondents perceive this factor as highly influential. However, 74\% of our respondents consider it as a low/moderate influential factor~\hist{icons/hreleasecyclefrequency.pdf}.

 \subsubsection{\noindent{\bf The quality attributes of the library.}} 
\label{subsubsec:quality}
Practitioners affirm that quality-related factors are strong determinants to decide whether to select a library. Given the existence of a large number of quality attributes, we introduce in this section the characteristics which were strongly emphasized in the interviews. These are \factor{documentation}, \factor{usability}, \factor{security vulnerabilities}, \factor{alignment with the architecture}, \factor{performance}, and \factor{test coverage}. 

\factor{Documentation} is, in general, perceived as a good sign of the library's quality. This factor is used in different ways among our participants. A group of our participants (P2, P3, P6, P7, P9, P10, P15, P16) examine the documentation up-front to determine if there are good examples or tutorials that clearly explain how to use the library. This is not necessarily the official documentation of the library but also information provided by forums or blog posts which describe examples of how the library is being used. The presence of this information helps practitioners to have a first impression of the library, in terms of how easy it is to use, and in general, what the library looks like. On the other hand, other participants (P1 and P5) emphasize the role of documentation during prototype development. For instance, P5 states that ``\textit{documentation is something that plays a role when you are trying to implement a proof of concept, not before}''. 51\% of our survey respondents perceive the quality of the documentation as a highly influential factor in the selection process, and 47\% consider it a low/moderate influential factor~\hist{icons/hqualitydocumentation.pdf}.   

Another quality characteristic extensively mentioned among participants (P2, P3, P7-P9, P11, P16) is \factor{usability} (i.e., how easy it is to use the library), which is seen as a positive characteristic in the selection process. For instance, P11 points out that ``\textit{if the library is very complex to use or causes a lot of overhead, I tend to go for another candidate}''. \factor{Usability} is identified as an influential factor by 55\% of our survey respondents who consider it a highly influential factor, and 42\% perceive it as low/moderate influential factor~\hist{icons/husability.pdf}.     

Practitioners (P5, P7, P9, P10, P13) also indicate that they tend to select libraries with good \factor{alignment between the library and the core technologies} used in their project or with a good match with the overall software architecture. P9 states that ``\textit{Compatibility with the software architecture is a factor that I check during the prototyping by identifying how easy it is to integrate the library to my project}''. Moreover, a considerable group of participants (P1, P2, P6, P7, P10, P12-P15, P16) highlights as well that the selection of software libraries is limited by the major technology choices made in the project. P12 mentions that ``\textit{If the core of our software is built using C++, we will use only libraries that support the language, we are not building wrappers to force the communication}''. 21\% of our survey respondents perceive the alignment between the library and the architectural decisions and technology stack of their projects as a highly influential factor in the library selection process, and 69\% consider it as a low/moderate influential factor~\hist{icons/halignment.pdf}.

\factor{How much a software library is tested} is perceived by some practitioners (P1, P10, P11, P15) as a good indicator of the quality of the library. Practitioners tend to discard libraries with low test coverage. Moreover, developers also take into account during the selection of libraries is knowing how often \factor{security vulnerabilities} occur (P1, P2, P5, P7, P11, P13, P15, P16). Interestingly, some participants affirm not to consider this factor up-front. For instance, P15 mentions that ``\textit{Checking security vulnerabilities is important, but I do not check it up-front. [I'd] probably using static analysis tools but they do not check everything}''. 
11\% of our survey respondents perceive that testing is a highly influential factor, 68\% of our respondents consider it a low/moderate influential factor, and 18\% believe this factor not to influence the selection process~\hist{icons/hwelltested.pdf}. Surprisingly, only 39\% of our respondents perceive the security aspects of the library as a highly influential factor, 53\% consider it a low/moderate influential factor, and only 4\% believe it does not play any role in the selection process~\hist{icons/hsecurity.pdf}.

Finally, we also observe that \factor{performance} is considered by practitioners (P2, P8-P9, P12, P15) as a critical quality attribute of a library. 
32\% of our survey respondents perceive that performance is a highly influential factor for the selection of a library, and 63\% consider it as a low/moderate influential factor~\hist{icons/hperformance.pdf}. 

\subsubsection{\noindent{\bf Functionalities provided by the library.}} 
\label{subsubsec:functionality}         

As expected, practitioners (P1-P5, P7, P8, P10, P13, P15-P16) tend to select libraries which offer a good match between their set of functionalities and the required features needed in the software. The \factor{fit for purpose} factor emerges as an essential criterion when exploring potential libraries~\hist{icons/hfitforpurpose.pdf}. 

Participants highlight that, besides \factor{fit for purpose}, it is crucial to consider the library's \factor{size and complexity}. In other words, the amount of code that the library has and whether the library offers way more functionalities than the ones needed. Practitioners perceive positively when the library provides the desired functionality with the least amount of code possible. For instance, P4 highlights that ``\textit{If 99\% of the library does something different then it is not a suitable candidate, so it does what you want, but you pull in too much junk}''. Participant P11 adds that ``\textit{If the library requires a lot of initialization code or even configuration, that can be a blocker in the selection}''. 16\% of our survey respondents perceive \factor{size and complexity} as a highly influential factor in the selection process~\hist{icons/hsizeandcomplexity.pdf}.

\subsubsection{\noindent{\bf Type of project.}}  
\label{subsubsec:typeproject}

How developers select libraries depends on the type of project of the software system under development, i.e., whether it is a \factor{brown- or green-field development}.

In the case of a \factor{green-field project}, the lack of restrictions enforced by existing software gives practitioners more freedom to make technological choices. For instance, participants (P1-P6, P11, P13, and P16) share the opinion that, for new projects, libraries are continuously adopted on-demand throughout the entire lifespan of a project. P5 highlights that ``\textit{If it is a new project, selecting libraries is more likely to happen because it might be a new setup or you still have all the freedom, so let's try this library or let's see if we can use it in this way}''. 

In the case of a \factor{brown-field project}, the selection of a software library is made upon a pre-existing software infrastructure. Our participants (P2, P4, P5, and P13) emphasize that, in this situation, decisions aim to find a suitable library aligned with the existing architecture. 
Interestingly, only 10\% of our survey respondents perceive the type of project as a highly influential factor. However, 18\% consider it not influential at all when selecting libraries~\hist{icons/hbrownorgreenfield.pdf}. 

\subsection{Human Factors} 
\label{sec:human-factors}

Human aspects play an essential role in practitioners' selection processes. Participants emphasize paying particular attention to the following factors: their individual perceptions, the community around the library, the different stakeholders of the project, and the characteristics of the company/organization where the library is being developed. 
\subsubsection{\noindent{\bf Individual.}}  
\label{subsubsec:individual}

The practitioners' personal experiences, perceptions, and knowledge about specific technologies influence their perception towards a software library. Practitioners (P3, P6, P10) mention that developers use their \factor{perception} when selecting libraries. For instance, P10 states: ``\textit{Our decisions are more based on emotions. What kind of feeling do we have towards a library?}'' However, P10 also points out that emotions can make such decisions difficult: ``\textit{When the decision is based on emotions or how do people feel about it, it is sometimes even harder to make people change their minds}''. 16\% of our survey respondents perceive personal feelings as a highly influential factor when selecting libraries, and 69\% consider it as a low/moderate influential factor~\hist{icons/hselfperception.pdf}. 

\subsubsection{\noindent{\bf Community.}} 
\label{subsubsec:community} 

Practitioners perceive that the community around a software library plays a positive and influential role in the selection process. Community is essential in two different contexts: the group of users who already adopted the library in their projects, and the group of developers who are maintaining and providing technical support for the library. These two groups offer useful input when new potential users are choosing libraries.

Practitioners (P1, P2, P4, P6, P10, P11, P13, P14, P16) consider the \factor{experience} that the community has with a library, in particular recommendations that other users of the library have (P1, P2, P4, P13) and what is the overall perception from the community towards the library (P6, P14). 
P16 points out that finding people that are tackling the same problem, and observing how they solved it with that library helps him reason about whether that library is the right choice: ``\textit{You can't be the first one having this problem. So when you search, you often encounter these libraries and often in web pages made not only by the company/developers creators of the library but also by the people who explain how they solved a certain problem using a certain library}''. P10 also wants to know how other companies (not an ``individual'') overcame problems that the library aims to solve; in other words, the experience that industry has with that library. 
23\% of our survey respondents perceive the collective experience of library users as a highly influential factor when selecting a library, and 63\% consider it as a low/moderate influential factor~\hist{icons/hcollectiveexperience.pdf}. 

Most participants (P1, P3-P6, P9-P16) consider the \factor{popularity} of a library to be an influential factor. We observe participants relying on different metrics, such as the number of stars (P3, P9, P11), the number of downloads (P9, P14), and the number of maven dependencies (P6). P6 emphasizes that ``\textit{I always looked to the number of maven dependencies which also indicates the importance of the library in its ecosystem. If lots of libraries use this one, it is a stable one. I think this might be an indication of stability in some way}''. Interestingly, P1 argues that popularity should also be seen through the evolution of the library in its different releases, i.e., whether the library is trending up or down. P1 says: ``\textit{If I check the library's trends in Google, I can see how frequently people look for it. This is a good indicator to see if the library's popularity is rising or not}''. 30\% of our survey respondents perceive the \textit{popularity} of a library as a highly influential factor in their selection process, and 59\% consider it as a low/moderate influential factor~\hist{icons/hpopularity.pdf}.


Besides the library's community seen from the perspective of its users, practitioners also pay attention to the characteristics of the contributors who are maintaining the library. Participants (P3, P7-P10, P14, P15) perceive an \factor{active community} supporting the library in a positive way. For instance, P15 emphasizes that high responsiveness to issues and questions on \textit{Stack Overflow} is a good indicator of an active community who supports the library. Some interviewees (P3, P7, P10, P12) also indicate that they consider it critical to know the number of contributors to the library, i.e., the size of the community behind the library. P10 points out that if the library has a significant impact on the project, it is also crucial to have people in the team to support the library if it gets deprecated: ``\textit{Does the library have a big impact in your project? [If] yes, then you need a team to support it. Otherwise, if the library is for some reason deprecated, that would represent a big risk. As a consequence, you get a piece of your application that you can not easily replace}''.   
30\% of our survey respondents perceive an active community supporting the library as a highly influential factor, and 62\% consider it as a low/moderate influential factor~\hist{icons/hactivecommunity.pdf}.

\subsubsection{\noindent{\bf Stakeholders.}} 
\label{subsubsec:stakeholders} 

Our interviews show that several stakeholders (i.e., the team, developers, project and product managers, external teams in the organization, and customers) might play a role in the selection process of software libraries. 

The selection of software libraries often starts with the initiative of a \textit{software developer}. Most of our interviewees (P1, P4-P16) agree that the choice of libraries is influenced by the opinions of the \factor{software development team}. 30\% of our survey respondents also perceive this factor as a highly influential factor, and 57\% consider it as a low/moderate influential factor~\hist{icons/hswdevteam.pdf}.  
We observe that having people in the team who faced a similar problem and/or had an experience using the library before saves time and facilitates the selection. 


However, software developers often do not have the authority or knowledge to decide which library to choose. Thus, the opinions formed by the software development team members move up in the chain. For example, we observe cases where a software architect or a technical lead had to approve a selection decision in the case of potential risks. In the words of P7: ``\textit{If it is a framework decision, the software architect or technical lead should be involved}''.

A few participants (P9, P12, P16) highlights that in some exceptional cases, \factor{the product manager and/or the project manager} might also influence the selection process. 
P16 exemplifies that, in some particular cases, the involvement of the \textit{project manager} is required: ``\textit{The project manager usually will be involved because when you use a library, there is also a certain kind of knowledge that you need about using the library. You have to learn how to incorporate it; you have to see whether it fits in your solution, etc. But, there are some costs involved in using the library which the software team might not be aware of. These costs might also influence the maintainability of the product}''. In contrast, only 3\% of our survey respondents perceive the roles of product project managers as highly influential factors in the selection of libraries. However, 36\% consider these roles do not influence the selection process~\hist{icons/hmanagers.pdf}.

Besides the software development team, practitioners (P1, P2, P8, P10, P15) perceive that \factor{other teams} within the organization might influence the final decision. For instance, P1 indicates that in the context of financial systems, new libraries should pass through a rigorous inspection before being used by software teams. In this case, a security team is often in charge of this inspection. They will check if the library is secure and meets all the requirements and policies of the organization. The organization might encourage software development teams to select libraries which belong to their software artifacts catalogue. However, if a new library is preferred, the team should provide strong arguments to support the requirement. In this case, the security team might have to inspect the library and finally approve its selection. 
Interestingly, only 4\% of our survey respondents consider \factor{other teams} as a highly influential factor, while 23\% consider this factor does not influence the selection process~\hist{icons/hotherteams.pdf}. 

Finally, practitioners (P2, P14, P16) perceive that \factor{customers} can also play an influential role in the library selection process in some cases. Customers regularly demand new features and services which have a significant impact on the software that provides those services. P16 exemplified ``\textit{Customers might have a large set of demands regarding security, privacy, etc. To incorporate a library within the project, we have to prove that the library meets all the requirements set by the customer}''. Additionally, P14 points out that customers might request to know how the library is managing customers' sensitive data. P14 states: ``\textit{We do business to business software. So there might be some customers who want to know how we handle the open-source projects regarding data protection}''. Interestingly, only 7\% of our survey respondents consider customers as a highly influential factor in the selection process, while 54\% perceive customers do not influence all~\hist{icons/hcustomers.pdf}.

\subsubsection{\noindent{\bf The organization.}}  

The organization (i.e., the company that the developers work for) itself plays a role in how software libraries are selected. The role of the management in technology decisions, the maturity of the process of library selection, the organization's culture and policies, and the type of industry are key factors that influence the process of library selection.

A few participants (P10, P11) have been in situations where the \factor{management and the strategy} of the company played an influential role in the selection of a library. P11 says: ``\textit{Sometimes the company makes strategic decisions and adopts a specific platform and that limits developers to use only compatible libraries or even a particular set of commercial libraries}''. However, only 8\% of our survey respondents perceive the management and the strategy of the company as a highly influential factor in the selection process, and 47\% consider it as a low/moderate influential factor. 35\% of our respondents believe it has no influence in the selection process at all~\hist{icons/hmanagement.pdf}.

The \factor{organizations' culture and policies} may also play an influential role when selecting libraries. We observe two different perspectives. In cases where the selection of libraries is not a fully-formalized process, practitioners (P1, P10, P14-P16) highlight that the organization might emphasize policies that software teams need to follow, e.g., policies regarding privacy and information protection. P10 states: ``\textit{Although we don't have a formal process for selecting libraries, we have very high standards for security. There are strict rules and inspections}''. In this case, a standard policy may include management of information, storage location, and time and security of the data transmission. 8\% of our survey respondents perceive this factor as a highly influential factor, and 52\% consider it as a low/moderate influential factor~\hist{icons/hcultureorganization.pdf}. However, other companies (as noted by P5, P10, P16) already have a mature process in place to select libraries, which naturally plays a role in the selection process. In the case of our interviewees, these mature processes are operationalized employing standard peer review processes (P5), strict procedures for security inspections while selecting libraries (P10), and following specific protocols for data protection (P16). P16 exemplifies ``\textit{In our company, for each project, we have one person responsible for everything related to information protection. So for each external component that we introduce to the project, it's investigated how this component affects our data privacy policies. It's a standard procedure}''.

Finally, the \factor{type of industry} (e.g., banking, oil and gas) that the software system under development belongs to plays an influential role when selecting libraries (P11, P14, P16). For instance, P14 highlights that in the case of business to business software development, customers need to be aware of how libraries manage sensitive information. On the other hand, only 9\% of our survey respondents consider that the \factor{type of industry} highly influences them when selecting libraries~~\hist{icons/htypeofindustry.pdf}. 

\subsection{Economic Factors}
\label{sec:economical-factors}

Software practitioners consider it crucial also to evaluate economic factors, such as total cost of ownership (TCO) and risks during the library selection process.    

\subsubsection{\noindent{\bf Total cost of ownership (TCO).}} 
\label{subsubsec:tco} 
The cost of owning a software library that is about to be selected is an important factor that our participants consider. For instance, the library's \factor{license}, the \textit{time and budget} available in case it is necessary to select a commercial library, the \textit{time} in case there is the possibility to develop the feature in-house instead of choosing a library, and the \textit{maintenance costs} associated with using a library. 

The \factor{license} of the software library is a crucial factor (P1, P2, P4, P5, P8-P10, P12-P16). P1 highlights how critical it is to have the rights to patch the library in case of bugs or security vulnerability issues: ``\textit{The license would influence my decision, but that would depend on the use case. Because if I need to apply a security fix, for example, it would be nice if we can change it ourselves. Especially if the library is not frequent in release cycles}''. Furthermore, P13 points out that the license sometimes is a business decision. It becomes an investment, especially when the decision is to select payware. P15 also suggests not to use a GPLv3 license if the aim is to develop commercial software\footnote{A GPLv3 license forces developers to open their software as well, a situation not desired in this type of business.}. Moreover, P16 emphasizes that it is vital to know if the customer plans to sell the solution as a product. If so, it is critical to guarantee that customers can incorporate all the source code under their name. In the case of proprietary libraries, it is sometimes a requirement to be able also to customize the code. Thus, the library's license should allow it. License is perceived as a highly influential factor by 45\% of our survey respondents, and 42\% consider it as a low/moderate influential factor~\hist{icons/hlicense.pdf}.  

The \factor{time and budget} available are also considered an essential factor. The possibility of choosing a commercial library is only an option if there is budget available (P9, P10, P14-P16). P10 exemplifies ``\textit{We will choose a proprietary license if it saves us time, risks, and we have money for it}''. Budget and time also play a role when the team decides to build the desired feature by themselves instead of choosing third-party libraries. P14 asks himself: ``\textit{How much does it cost to build the desired functionality by ourselves instead of using a library?}'' 39\% of our survey respondents perceive this factor as a highly influential factor in the selection process, and 48\% consider it as a low/moderate influential factor~\hist{icons/htimebudget.pdf}. 

\subsubsection{\noindent{\bf Risk.}}
\label{subsubsec:risk}  

The possible risks that a library might introduce to the software system, primarily affecting budget and development time, is also an essential factor that practitioners take into consideration (P2, P4, P10, P11, P14), also perceived as a highly influential factor by 38\% of our survey respondents. Only 5\% consider the risk to have no influence on the selection process~\hist{icons/hrisks.pdf}.

For example, P2 says that, at her/his company, basic risk assessments include the time needed to integrate the third-party code into the project, and an estimate of how long the library would be available in the future: ``\textit{How much development time is it going to cost me to integrate it in my project? And then also, how likely is the library to stay available in the future?}''. P10 also exemplifies ``\textit{By using a library ready-to-use, proven, I'm avoiding risks and reducing the time for its integration to the project}''. P14 mentions that, in her/his case, libraries that are already integrated into the main framework of her/his company are less risky, and therefore, more likely to be selected.

The presence of an organization behind the library provides practitioners (P1, P5, P8-P10, P14, P16) with a good perception of trustworthiness of the library. Indeed, participants associate the existence of a supporting organization with lower risk. For instance, P16 emphasizes that ``\textit{My decision would be influenced if there is a company behind the library. For example, there are libraries where Microsoft provides help as a major stakeholder so I'll use them}''.


\subsection{The role of the factors in the selection process}
\label{sec:interplay}

Throughout the interviews, participants explained not only which factors they take into consideration, but also in ``what moment'' that happens. And while we observe that these factors can emerge at any time during the process, some factors are often taken into account more at some stages than others. In this section, we map the factors mentioned above with the stage in which they become relevant (which we also show in Table~\ref{tab:all-factors}, column U/P).

As expected, most of our participants (P1, P4, P6, P7, P9-P12, P15, P16) highlight that they follow an ad-hoc approach unless organizations introduce strict policies in the adoption of third-party code in their software projects (P2, P7, P10, P15). Nevertheless, we are able to generalize their decision process into two main steps: (i) the up-front selection, and (ii) the prototyping stage. The up-front selection step happens when developers are researching the existing libraries that are out there. Libraries that are positively evaluated at this step become candidate libraries; others are discarded. Developers take the candidate libraries which they then take to the next step: the prototype stage. In this step, practitioners evaluate the libraries by developing proofs-of-concept. The best candidate out of the prototyping stage is then selected.\footnote{While our description makes the selection process to resemble a linear process, in practice, developers perform several iterations, i.e., research, find candidates, prototype, repeat until they are satisfied with the chosen library.}

Regarding the boundaries of the decision space, according to our participants, the first factor taken into account is the \factor{type of project}. Our interviewees point out that green-field development projects provide them with absolute freedom of choice. According to them, at this point, given that the first technology choices are still occurring (e.g. frameworks, programming language, databases, etc.), practitioners have more freedom to experiment and try out new libraries. On the contrary, others also state that brown-field development introduces constraints imposed by the dominant architectural framework or its major components. Thus, any library selected should fit the existing technological stack of the project. 

At the \textit{up-front} stage, practitioners highlight that most of the human factors influence them when selecting libraries. However, at this point, these factors emerge as strong determinants: \factor{practitioners' individual experience and perception} of the library, the \factor{community experience} of the library users, the \factor{popularity} of the library, and how \factor{responsive} the community is in providing support and fixing bugs. Concerning technical aspects, practitioners also point out that, at this stage, they pay more attention to the \factor{maturity} and \factor{stability}, the quality of the \factor{documentation}, and whether the library is \factor{actively maintained}. Additionally, regarding the economic characteristics of the library, practitioners emphasize that at this stage, the type of \factor{license} plays a critical role in the selection process, followed by \factor{time and budget}.           

At \textit{prototyping} stage, human factors also play a critical role. If practitioners do not have enough experience or familiarity with the technologies associated with the library, they tend to gather the \factor{knowledge of their immediate community} (e.g. teammates or colleagues), or \factor{feedback from external users} as well. Concerning technical aspects of the library, practitioners take \factor{usability} and the \factor{quality of the documentation} more into account. Indeed, participants note that, while usability and documentation can be somewhat evaluated at the up-front stage, they better experience them during the prototyping stage. At this stage, practitioners also tend to evaluate the \factor{impact of the library on the overall architecture}. Common criteria used at this level are \factor{performance} and \factor{security}. It is worth mentioning that economic factors play a significant role at this stage. After performing a \factor{risk assessment} of a candidate library, and depending on the \factor{time and budget} available in the organization, the decision space can include the selection of a paid version of the library, the development of the required feature by themselves, or the outsourcing of the development of the feature.

\section{Related Work}
\label{sec:body-of-knowledge}

\begin{table}
\footnotesize
\begin{tabular}{p{2.8cm}p{5cm}}
\toprule
\textbf{Paper} & \textbf{Factors} \\
\midrule 
de la Mora and Nadi~\cite{delamora2018b, delamora2018a} & popularity, release frequency, issue response time and issue closing time, backwards compatibility, performance and security, last modification date, last discussed on stack overflow, documentation, functionality, easy of use \\
Pano et al.~\cite{pano2016} & performance, size, automatization, learnability, complexity, understandability, competitor analysis, collegial advice, community size, community responsiveness, suitability, updates, modularity, isolation, extensibility, cost, customer, developer, team, team leader \\
Xu et al.~\cite{xu2020} & libraries well maintained, well tested, meets the requirements, actively developed, update cadence, popularity, respected developers, widely adopted, used by other projects, exposure on Stack Overflow, documentation readability, clearness of the code, stability, size or complexity, license compatibility, ease of integration\\
Hora and Valente~\cite{hora2015} & popularity \\
Abdalkareem et al.~\cite{abdalkareem2017} & well implemented and tested, well maintained code, readability, performance, security, license compatibility \\
Piccioni et al.~\cite{piccioni2013} & usability, documentation\\
Myers and Stylos~\cite{myers2016} & usability (learnability, simplicity, consistency)\\
Gizas et al.~\cite{gizas2012} & performance \\
\bottomrule
\end{tabular}
\caption{Key factors in the process of selecting software libraries that have emerged in the previous literature.}
\label{tab:factors-related-work}
\vspace{-9mm}
\end{table}

In Table~\ref{tab:factors-related-work}, we show the key factors that emerged in related research on the selection process of software libraries. Some of the factors that we observe in our work are inline with the existing current body of knowledge. 

More specifically, the factors associated with \textit{usability} (e.g., learnability, size and complexity, understandability, readability, simplicity, documentation, ease of integration, etc.), \textit{performance}, and \textit{popularity} have been highlighted in different studies~\cite{delamora2018a,delamora2018b,pano2016,xu2020,hora2015,abdalkareem2017,piccioni2013,myers2016} as the most influential factors when practitioners select libraries. The literature also emphasizes the significant role of factors associated with the \textit{maintainability} of the library (e.g., release frequency, issue response time and issue closing time, last modification date, community size), how well the software library is \textit{tested}, as well as its \textit{license} compatibility.

Our study reveals the importance of factors that were, so far, not known to our community (in Table~\ref{tab:all-factors}, we mark all the new factors with a checkmark). 
Concerning technical factors, our study highlights the role of the \factor{type of project} of the software system under development. Being a software project in the context of a green-field or brown-field development affects the freedom of choice (i.e., the decision space). Concerning human factors, another aspect not known to current literature is the influential role of some stakeholders within the organization (e.g., product and project managers, and other teams). In the context of organizational factors, practitioners perceive the type of industry, as well as the company's culture and policies and management and strategy as influential factors. The developer's self-perception towards a library also seems to influence the selection process. Finally, regarding \textit{economic factors}, unlike other studies, practitioners also pointed out the role that a risk assessment plays in the decision when selecting a library.

Interestingly, for most of the new factors we introduce, mostly regarding the role of organization and stakeholders to the process, we observe that survey respondents did not fully perceive them as highly influential. The role of customers, for example, is considered to be not influential by 54\% of the survey respondents. Or, the type of project (being a brown- or a green-field project), while survey respondents perceive it to play a low/moderate role in the process, only 10\% believe it has a high influence. We argue that this indicates the importance of the context, i.e., industry, where the decision is made. Our interviewees work for various types of industries, ranging from banking, gas and oil, to consulting and research. 

When looking at the related work, we observe participants coming from different backgrounds; however, the type of industry that they work in is not clear. As examples, in work from de la Mora and Nadi~\cite{delamora2018a, delamora2018b}, participants come from GitHub and Stack Overflow (besides 49\% of them being students); in the work of Pano et al.~\cite{pano2016}, participants come from different company sizes (ranging from small to large), but there is no indication as to which industry they are in. We, therefore, conjecture that the types of industry that we studied explain the new factors we found as well as that they did not generalize well in the survey. Future research should understand more deeply the role of industry in the selection process of software libraries.

We also see works that focus on specific programming languages or software ecosystems (e.g. JavaScript~\cite{pano2016}, Android and Python~\cite{xu2020}, Java~\cite{hora2015}, Nodejs~\cite{abdalkareem2017}). The work of Pano et al.~\cite{pano2016}, which we discuss above, focuses solely on JavaScript frameworks (frameworks only, not all types of possible software libraries). We did not control for a programming language or software ecosystem or type of library in our work. Such a design decision in our study enables us to devise a broader and more generic set of factors. However, we agree that similar to the industry context; different programming languages might have specific factors, which should be explored in future work.

Finally, we observed in the literature that the Unified Theory of Acceptance and Use of Technology (UTAUT) proposed by Venkatesh et al.~\cite{venkatesh2012} has been extensively adopted to explain the acceptance of technology by users. They highlight seven constructs as direct determinants of behavioral intention to use technology, namely performance expectancy, effort expectancy, social influence, facilitating conditions, hedonic motivation, price value, and habit. We note that the \numberoffactors factors, highlighted in our study, categorized into technical, human and economic factors are aligned with a UTAUT category and reinforce our contribution to the existing literature, especially regarding the influential role of organizational aspects, stakeholders experience and users' self-perception.   

\section{Implications}
\label{sec-discussion}

In this section, we discuss the implications of our work to library maintainers, library users, package manager developers, and researchers.

\cooltitle{Implications to library maintainers.}
We observe that practitioners often discard libraries with poor documentation, low maintenance activity and/or no long-term support plan, no stable releases, security vulnerabilities reported but not fixed quickly, and negative reputation in the community. Interestingly, our findings match with suggestions provided by Pano et al.~\cite{pano2016} in their study regarding factors and actors leading to the adoption of a JavaScript framework.  

\cooltitle{Implications to potential library users.}
Our work shows that the number of factors that may play a role in the selection process of a library is substantial. Software teams following an ad-hoc process might miss some of them, and a wrongly selected library might lead to higher maintenance costs and other issues in the future.

Based on the factors we identified, a systematic evaluation of a library should include the following reasoning: Do I have a positive experience using this library? What is the overall perception of the library among its users? Is it a popular library? How active is the community behind the library providing support and fixing bugs? How mature and stable is the library? Does the library have proper documentation? Is the library well-maintained? Does the library have an appropriate license for my project? Do budget and time impose any constraints? 

We believe that, by taking into account these considerations, practitioners would be in a better position to make systematic decisions while selecting libraries.

\cooltitle{Implications to package manager developers and the need for modern package managers.} Practitioners do not have access to appropriate unified infrastructure to support the selection process of software libraries. To tackle this situation, our work points to a possible direction: reinforce the role of package managers in the ecosystem, turning package managers into a single point of decision.

Most of the metadata that current package managers offer is focused on library popularity. Research has shown that having information available about the elements which affect API popularity can be beneficial when selecting a library~\cite{mileva2010}. Our work reveals several other vital quantitative factors that developers usually take into account in the decision-making process. For instance, maintenance activity, release frequency, community size, and quality aspects of the library are among the most critical factors to identify if the library is getting momentum. However, in diverse ecosystems, this information is currently dispersed and not readily available for decision-makers.    

The orchestration role of package managers situates them in an excellent position to provide library users with the right information when selecting a library (single point of decision). We, therefore, suggest that \textit{modern package managers} should be designed to facilitate the selection process of libraries as well. Through appropriate library metadata aggregation from multiple repositories, package managers would take the leading role among all information sources for exploring software libraries.


\cooltitle{Implications to researchers.} The factors we have catalogued in this work complement the ones identified in related work, and they essentially represent information needs that developers have during library selection processes. Our study also exposes the need for appropriate (near) real-time data-driven feedback among practitioners. The rapid evolution and overgrowing number of third-party libraries push forward the need to investigate and develop innovative tools aligned with these information needs. For example, we have identified risk assessment as one of the factors involved in library selection decisions---how can we measure such risk and possibly develop tools to enable automated risk assessment?

Many of the factors that our study uncovered and that had not been identified by previous work were \textit{human factors}. While our work provides a first categorization of these factors, we believe future work is necessary to solidify our understanding of how human factors impact library selection. In particular, given the wide range of stakeholders involved, reconciling potentially conflicting views on the same library selection decision is a significant challenge which requires further investigation.

\subsection{Threats to validity}
\label{sec:threats-to-validity}

In the following, we address the validity of this study in the context of qualitative research~\cite{guba1981criteria,korstjens2018series}. 

\textit{Transferability.} 
Transferability is the degree to which our results can be transferred to other contexts. Our study is based on semi-structured interviews collecting the experience of 16 participants and a validation survey. Given that their experiences, companies, technology stacks, and business domains varied considerably, the factors identified in our study should fit most software development companies.  
We did not include open source developers in the study, and therefore, we suggest future research to understand whether that community considers the factors identified.

\textit{Credibility.} 
Credibility is about whether the research findings are correctly drawn from the original data. We applied four different strategies to ensure credibility:
\begin{enumerate*}[label=(\alph*)] 
    \item the list of factors was iteratively developed by two researchers (the first two authors of this paper)---it is worth noting that a single researcher performed open coding of the transcribed interview data and to ensure quality, the researcher listened to each recording two times;
    \item the set of factors and findings were discussed several times between all the authors of this paper to mitigate bias from the researchers involved in the study;
    \item the factors were corroborated through six-member checking sessions; and 
    \item we challenged our qualitative findings by conducting a survey with \numberofsurveys participants. 
\end{enumerate*}

\textit{Confirmability.} 
Confirmability is the degree to which other researchers can confirm the findings. We do not have the participants' (and their employers') permission to share the transcription of the interviews. We tried as much as possible to show evidence for each factor by quoting participants. We make our interview script, survey form and survey answers available in our online appendix~\cite{appendix}.

\section{Conclusion}
\label{sec-conclusion}

The software industry relies heavily on the reuse of third-party libraries. Choosing the best library is, however, a demanding task that will only become more complicated, given the ever-growing complexity of software systems and the number of libraries available. In this paper, we systematically documented a comprehensive set of \numberoffactors factors for library selection done by practitioners, including \numberofnewfactors factors not previously included in literature. With these factors documented, practitioners can now start adopting more systematic approaches to third-party library selection. We foresee the creation of tools to better integrate prominent factors into the library selection process, in particular for technical factors such as quality and release factors, as well as for human and economic factors like community activeness and popularity, risk, and cost of ownership.
\section*{Acknowledgement}

We would like to thank the 16 interviewees for their availability in this study and the 115 software practitioners who completed the online survey. 
This work was partially funded by NWO grant 628.008.001 (CodeFeedr), H2020 grant 825328 (FASTEN), and ARC grant DE180100153.

\bibliographystyle{ACM-Reference-Format}
\bibliography{refs}


\begin{thebibliography}{29}


\ifx \showCODEN    \undefined \def \showCODEN     #1{\unskip}     \fi
\ifx \showDOI      \undefined \def \showDOI       #1{#1}\fi
\ifx \showISBNx    \undefined \def \showISBNx     #1{\unskip}     \fi
\ifx \showISBNxiii \undefined \def \showISBNxiii  #1{\unskip}     \fi
\ifx \showISSN     \undefined \def \showISSN      #1{\unskip}     \fi
\ifx \showLCCN     \undefined \def \showLCCN      #1{\unskip}     \fi
\ifx \shownote     \undefined \def \shownote      #1{#1}          \fi
\ifx \showarticletitle \undefined \def \showarticletitle #1{#1}   \fi
\ifx \showURL      \undefined \def \showURL       {\relax}        \fi
\providecommand\bibfield[2]{#2}
\providecommand\bibinfo[2]{#2}
\providecommand\natexlab[1]{#1}
\providecommand\showeprint[2][]{arXiv:#2}

\bibitem[\protect\citeauthoryear{Abdalkareem, Nourry, Wehaibi, Mujahid, and
  Shihab}{Abdalkareem et~al\mbox{.}}{2017}]%
        {abdalkareem2017}
\bibfield{author}{\bibinfo{person}{Rabe Abdalkareem}, \bibinfo{person}{Olivier
  Nourry}, \bibinfo{person}{Sultan Wehaibi}, \bibinfo{person}{Suhaib Mujahid},
  {and} \bibinfo{person}{Emad Shihab}.} \bibinfo{year}{2017}\natexlab{}.
\newblock \showarticletitle{Why Do Developers Use Trivial Packages? An
  Empirical Case Study on Npm}. In \bibinfo{booktitle}{\emph{Proceedings of the
  2017 11th Joint Meeting on Foundations of Software Engineering}} (Paderborn,
  Germany) \emph{(\bibinfo{series}{ESEC/FSE 2017})}.
  \bibinfo{publisher}{Association for Computing Machinery},
  \bibinfo{address}{New York, NY, USA}, \bibinfo{pages}{385–395}.
\newblock
\showISBNx{9781450351058}
\urldef\tempurl%
\url{https://doi.org/10.1145/3106237.3106267}
\showDOI{\tempurl}


\bibitem[\protect\citeauthoryear{Arnott}{Arnott}{2005}]%
        {arnott2005}
\bibfield{author}{\bibinfo{person}{David Arnott}.}
  \bibinfo{year}{2005}\natexlab{}.
\newblock \showarticletitle{Cognitive biases and decision support systems
  development: a design science approach}.
\newblock \bibinfo{journal}{\emph{Information Systems Journal}}
  \bibinfo{volume}{16}, \bibinfo{number}{1} (\bibinfo{date}{2018/09/28}
  \bibinfo{year}{2005}), \bibinfo{pages}{55--78}.
\newblock
\showISBNx{1350-1917}
\urldef\tempurl%
\url{https://doi.org/10.1111/j.1365-2575.2006.00208.x}
\showDOI{\tempurl}


\bibitem[\protect\citeauthoryear{Creswell}{Creswell}{2013}]%
        {creswell2013research}
\bibfield{author}{\bibinfo{person}{John~W Creswell}.}
  \bibinfo{year}{2013}\natexlab{}.
\newblock \bibinfo{booktitle}{\emph{Research design: Qualitative, quantitative,
  and mixed methods approaches}}.
\newblock \bibinfo{publisher}{Sage publications}.
\newblock


\bibitem[\protect\citeauthoryear{de~la Mora and Nadi}{de~la Mora and
  Nadi}{2018a}]%
        {delamora2018b}
\bibfield{author}{\bibinfo{person}{Fernando~L\'{o}pez de~la Mora} {and}
  \bibinfo{person}{Sarah Nadi}.} \bibinfo{year}{2018}\natexlab{a}.
\newblock \showarticletitle{An Empirical Study of Metric-based Comparisons of
  Software Libraries}. In \bibinfo{booktitle}{\emph{Proceedings of the 14th
  International Conference on Predictive Models and Data Analytics in Software
  Engineering}} (Oulu, Finland) \emph{(\bibinfo{series}{PROMISE'18})}.
  \bibinfo{publisher}{ACM}, \bibinfo{address}{New York, NY, USA},
  \bibinfo{pages}{22--31}.
\newblock
\showISBNx{978-1-4503-6593-2}
\urldef\tempurl%
\url{https://doi.org/10.1145/3273934.3273937}
\showDOI{\tempurl}


\bibitem[\protect\citeauthoryear{de~la Mora and Nadi}{de~la Mora and
  Nadi}{2018b}]%
        {delamora2018a}
\bibfield{author}{\bibinfo{person}{Fernando~L\'{o}pez de~la Mora} {and}
  \bibinfo{person}{Sarah Nadi}.} \bibinfo{year}{2018}\natexlab{b}.
\newblock \showarticletitle{Which Library Should I Use?: A Metric-based
  Comparison of Software Libraries}. In \bibinfo{booktitle}{\emph{Proceedings
  of the 40th International Conference on Software Engineering: New Ideas and
  Emerging Results}} (Gothenburg, Sweden) \emph{(\bibinfo{series}{ICSE-NIER
  '18})}. \bibinfo{publisher}{ACM}, \bibinfo{address}{New York, NY, USA},
  \bibinfo{pages}{37--40}.
\newblock
\showISBNx{978-1-4503-5662-6}
\urldef\tempurl%
\url{https://doi.org/10.1145/3183399.3183418}
\showDOI{\tempurl}


\bibitem[\protect\citeauthoryear{Gizas, Christodoulou, and Papatheodorou}{Gizas
  et~al\mbox{.}}{2012}]%
        {gizas2012}
\bibfield{author}{\bibinfo{person}{Andreas Gizas}, \bibinfo{person}{Sotiris
  Christodoulou}, {and} \bibinfo{person}{Theodore Papatheodorou}.}
  \bibinfo{year}{2012}\natexlab{}.
\newblock \showarticletitle{Comparative Evaluation of Javascript Frameworks}.
  In \bibinfo{booktitle}{\emph{Proceedings of the 21st International Conference
  on World Wide Web}} (Lyon, France) \emph{(\bibinfo{series}{WWW ’12
  Companion})}. \bibinfo{publisher}{Association for Computing Machinery},
  \bibinfo{address}{New York, NY, USA}, \bibinfo{pages}{513–514}.
\newblock
\showISBNx{9781450312301}
\urldef\tempurl%
\url{https://doi.org/10.1145/2187980.2188103}
\showDOI{\tempurl}


\bibitem[\protect\citeauthoryear{Guba}{Guba}{1981}]%
        {guba1981criteria}
\bibfield{author}{\bibinfo{person}{Egon~G Guba}.}
  \bibinfo{year}{1981}\natexlab{}.
\newblock \showarticletitle{Criteria for assessing the trustworthiness of
  naturalistic inquiries}.
\newblock \bibinfo{journal}{\emph{Educational Technology research and
  development}} \bibinfo{volume}{29}, \bibinfo{number}{2}
  (\bibinfo{year}{1981}), \bibinfo{pages}{75--91}.
\newblock


\bibitem[\protect\citeauthoryear{Hiller and Diluzio}{Hiller and
  Diluzio}{2004}]%
        {hiller2004}
\bibfield{author}{\bibinfo{person}{Harry~H. Hiller} {and}
  \bibinfo{person}{Linda Diluzio}.} \bibinfo{year}{2004}\natexlab{}.
\newblock \showarticletitle{The Interviewee and the Research Interview:
  Analysing a Neglected Dimension in Research*}.
\newblock \bibinfo{journal}{\emph{Canadian Review of Sociology/Revue canadienne
  de sociologie}} \bibinfo{volume}{41}, \bibinfo{number}{1}
  (\bibinfo{year}{2004}), \bibinfo{pages}{1--26}.
\newblock
\urldef\tempurl%
\url{https://doi.org/10.1111/j.1755-618X.2004.tb02167.x}
\showDOI{\tempurl}


\bibitem[\protect\citeauthoryear{Hora and Valente}{Hora and Valente}{2015}]%
        {hora2015}
\bibfield{author}{\bibinfo{person}{A. Hora} {and} \bibinfo{person}{M.~T.
  Valente}.} \bibinfo{year}{2015}\natexlab{}.
\newblock \showarticletitle{Apiwave: Keeping track of API popularity and
  migration}. In \bibinfo{booktitle}{\emph{2015 IEEE International Conference
  on Software Maintenance and Evolution (ICSME)}}, Vol.~\bibinfo{volume}{00}.
  \bibinfo{pages}{321--323}.
\newblock
\urldef\tempurl%
\url{https://doi.org/10.1109/ICSM.2015.7332478}
\showDOI{\tempurl}


\bibitem[\protect\citeauthoryear{Hove and Anda}{Hove and Anda}{2005}]%
        {hove2005experiences}
\bibfield{author}{\bibinfo{person}{Siw~Elisabeth Hove} {and}
  \bibinfo{person}{Bente Anda}.} \bibinfo{year}{2005}\natexlab{}.
\newblock \showarticletitle{Experiences from conducting semi-structured
  interviews in empirical software engineering research}. In
  \bibinfo{booktitle}{\emph{Software metrics, 2005. 11th ieee international
  symposium}}. IEEE, \bibinfo{pages}{10--pp}.
\newblock


\bibitem[\protect\citeauthoryear{Korstjens and Moser}{Korstjens and
  Moser}{2018}]%
        {korstjens2018series}
\bibfield{author}{\bibinfo{person}{Irene Korstjens} {and}
  \bibinfo{person}{Albine Moser}.} \bibinfo{year}{2018}\natexlab{}.
\newblock \showarticletitle{Series: Practical guidance to qualitative research.
  Part 4: Trustworthiness and publishing}.
\newblock \bibinfo{journal}{\emph{European Journal of General Practice}}
  \bibinfo{volume}{24}, \bibinfo{number}{1} (\bibinfo{year}{2018}),
  \bibinfo{pages}{120--124}.
\newblock


\bibitem[\protect\citeauthoryear{Kula, German, Ishio, and Inoue}{Kula
  et~al\mbox{.}}{2015}]%
        {kula2015}
\bibfield{author}{\bibinfo{person}{R.~G. Kula}, \bibinfo{person}{D.~M. German},
  \bibinfo{person}{T. Ishio}, {and} \bibinfo{person}{K. Inoue}.}
  \bibinfo{year}{2015}\natexlab{}.
\newblock \showarticletitle{Trusting a library: A study of the latency to adopt
  the latest Maven release}, In \bibinfo{booktitle}{2015 IEEE 22nd
  International Conference on Software Analysis, Evolution, and Reengineering
  (SANER)}.
\newblock \bibinfo{journal}{\emph{2015 IEEE 22nd International Conference on
  Software Analysis, Evolution, and Reengineering (SANER)}},
  \bibinfo{pages}{520--524}.
\newblock
\showISBNx{1534-5351}
\urldef\tempurl%
\url{https://doi.org/10.1109/SANER.2015.7081869}
\showDOI{\tempurl}


\bibitem[\protect\citeauthoryear{Larios~Vargas, Aniche, Treude, Bruntink, and
  Gousios}{Larios~Vargas et~al\mbox{.}}{2020}]%
        {appendix}
\bibfield{author}{\bibinfo{person}{Enrique Larios~Vargas},
  \bibinfo{person}{Maurício Aniche}, \bibinfo{person}{Christoph Treude},
  \bibinfo{person}{Magiel Bruntink}, {and} \bibinfo{person}{Georgios Gousios}.}
  \bibinfo{year}{2020}\natexlab{}.
\newblock \bibinfo{booktitle}{\emph{{Selecting third-party libraries: The
  practitioners' perspective}}}.
\newblock
\urldef\tempurl%
\url{https://doi.org/10.5281/zenodo.3979446}
\showDOI{\tempurl}


\bibitem[\protect\citeauthoryear{Larios~Vargas, Hejderup, Kechagia, Bruntink,
  and Gousios}{Larios~Vargas et~al\mbox{.}}{2018}]%
        {larios2018}
\bibfield{author}{\bibinfo{person}{Enrique Larios~Vargas},
  \bibinfo{person}{Joseph Hejderup}, \bibinfo{person}{Maria Kechagia},
  \bibinfo{person}{Magiel Bruntink}, {and} \bibinfo{person}{Georgios Gousios}.}
  \bibinfo{year}{2018}\natexlab{}.
\newblock \showarticletitle{Enabling Real-time Feedback in Software
  Engineering}. In \bibinfo{booktitle}{\emph{Proceedings of the 40th
  International Conference on Software Engineering: New Ideas and Emerging
  Results}} (Gothenburg, Sweden) \emph{(\bibinfo{series}{ICSE-NIER '18})}.
  \bibinfo{publisher}{ACM}, \bibinfo{address}{New York, NY, USA},
  \bibinfo{pages}{21--24}.
\newblock
\showISBNx{978-1-4503-5662-6}
\urldef\tempurl%
\url{https://doi.org/10.1145/3183399.3183416}
\showDOI{\tempurl}


\bibitem[\protect\citeauthoryear{{Li}, {Wang}, {Wang}, {Wang}, {Wu}, {Liu},
  {Xue}, and {Huo}}{{Li} et~al\mbox{.}}{2017}]%
        {li2017}
\bibfield{author}{\bibinfo{person}{M. {Li}}, \bibinfo{person}{W. {Wang}},
  \bibinfo{person}{P. {Wang}}, \bibinfo{person}{S. {Wang}}, \bibinfo{person}{D.
  {Wu}}, \bibinfo{person}{J. {Liu}}, \bibinfo{person}{R. {Xue}}, {and}
  \bibinfo{person}{W. {Huo}}.} \bibinfo{year}{2017}\natexlab{}.
\newblock \showarticletitle{LibD: Scalable and Precise Third-Party Library
  Detection in Android Markets}. In \bibinfo{booktitle}{\emph{2017 IEEE/ACM
  39th International Conference on Software Engineering (ICSE)}}.
  \bibinfo{pages}{335--346}.
\newblock
\showISSN{1558-1225}
\urldef\tempurl%
\url{https://doi.org/10.1109/ICSE.2017.38}
\showDOI{\tempurl}


\bibitem[\protect\citeauthoryear{Lima and Hora}{Lima and Hora}{2019}]%
        {lima2019}
\bibfield{author}{\bibinfo{person}{Caroline Lima} {and} \bibinfo{person}{Andre
  Hora}.} \bibinfo{year}{2019}\natexlab{}.
\newblock \showarticletitle{What are the characteristics of popular APIs? A
  large-scale study on Java, Android, and 165 libraries}.
\newblock \bibinfo{journal}{\emph{Software Quality Journal}}
  (\bibinfo{year}{2019}).
\newblock
\showISBNx{1573-1367}
\urldef\tempurl%
\url{https://doi.org/10.1007/s11219-019-09476-z}
\showDOI{\tempurl}


\bibitem[\protect\citeauthoryear{Mileva, Dallmeier, Burger, and Zeller}{Mileva
  et~al\mbox{.}}{2009}]%
        {mileva2009}
\bibfield{author}{\bibinfo{person}{Yana~Momchilova Mileva},
  \bibinfo{person}{Valentin Dallmeier}, \bibinfo{person}{Martin Burger}, {and}
  \bibinfo{person}{Andreas Zeller}.} \bibinfo{year}{2009}\natexlab{}.
\newblock \showarticletitle{Mining Trends of Library Usage}. In
  \bibinfo{booktitle}{\emph{Proceedings of the Joint International and Annual
  ERCIM Workshops on Principles of Software Evolution (IWPSE) and Software
  Evolution (Evol) Workshops}} (Amsterdam, The Netherlands)
  \emph{(\bibinfo{series}{IWPSE-Evol '09})}. \bibinfo{publisher}{ACM},
  \bibinfo{address}{New York, NY, USA}, \bibinfo{pages}{57--62}.
\newblock
\showISBNx{978-1-60558-678-6}
\urldef\tempurl%
\url{https://doi.org/10.1145/1595808.1595821}
\showDOI{\tempurl}


\bibitem[\protect\citeauthoryear{Mileva, Dallmeier, and Zeller}{Mileva
  et~al\mbox{.}}{2010}]%
        {mileva2010}
\bibfield{author}{\bibinfo{person}{Yana~Momchilova Mileva},
  \bibinfo{person}{Valentin Dallmeier}, {and} \bibinfo{person}{Andreas
  Zeller}.} \bibinfo{year}{2010}\natexlab{}.
\newblock \showarticletitle{Mining API Popularity}. In
  \bibinfo{booktitle}{\emph{Testing --Practice and Research Techniques}},
  \bibfield{editor}{\bibinfo{person}{Leonardo Bottaci} {and}
  \bibinfo{person}{Gordon Fraser}} (Eds.). \bibinfo{publisher}{Springer Berlin
  Heidelberg}, \bibinfo{address}{Berlin, Heidelberg},
  \bibinfo{pages}{173--180}.
\newblock
\showISBNx{978-3-642-15585-7}


\bibitem[\protect\citeauthoryear{Milkman, Chugh, and Bazerman}{Milkman
  et~al\mbox{.}}{2009}]%
        {milkman2009}
\bibfield{author}{\bibinfo{person}{Katherine~L. Milkman},
  \bibinfo{person}{Dolly Chugh}, {and} \bibinfo{person}{Max~H. Bazerman}.}
  \bibinfo{year}{2009}\natexlab{}.
\newblock \showarticletitle{How Can Decision Making Be Improved?}
\newblock \bibinfo{journal}{\emph{Perspectives on psychological science : a
  journal of the Association for Psychological Science}}  \bibinfo{volume}{4 4}
  (\bibinfo{year}{2009}), \bibinfo{pages}{379--83}.
\newblock


\bibitem[\protect\citeauthoryear{Mojica, Adams, Nagappan, Dienst, Berger, and
  Hassan}{Mojica et~al\mbox{.}}{2014}]%
        {mojica2014}
\bibfield{author}{\bibinfo{person}{I.~J. Mojica}, \bibinfo{person}{B. Adams},
  \bibinfo{person}{M. Nagappan}, \bibinfo{person}{S. Dienst},
  \bibinfo{person}{T. Berger}, {and} \bibinfo{person}{A.~E. Hassan}.}
  \bibinfo{year}{2014}\natexlab{}.
\newblock \showarticletitle{A Large-Scale Empirical Study on Software Reuse in
  Mobile Apps}.
\newblock \bibinfo{journal}{\emph{IEEE Software}} \bibinfo{volume}{31},
  \bibinfo{number}{2} (\bibinfo{date}{Mar} \bibinfo{year}{2014}),
  \bibinfo{pages}{78--86}.
\newblock
\showISSN{0740-7459}
\urldef\tempurl%
\url{https://doi.org/10.1109/MS.2013.142}
\showDOI{\tempurl}


\bibitem[\protect\citeauthoryear{Myers and Stylos}{Myers and Stylos}{2016}]%
        {myers2016}
\bibfield{author}{\bibinfo{person}{Brad~A. Myers} {and}
  \bibinfo{person}{Jeffrey Stylos}.} \bibinfo{year}{2016}\natexlab{}.
\newblock \showarticletitle{Improving API Usability}.
\newblock \bibinfo{journal}{\emph{Commun. ACM}} \bibinfo{volume}{59},
  \bibinfo{number}{6} (\bibinfo{date}{May} \bibinfo{year}{2016}),
  \bibinfo{pages}{62–69}.
\newblock
\showISSN{0001-0782}
\urldef\tempurl%
\url{https://doi.org/10.1145/2896587}
\showDOI{\tempurl}


\bibitem[\protect\citeauthoryear{Nguyen, Rocco, Ruscio, and Penta}{Nguyen
  et~al\mbox{.}}{2020}]%
        {nguyen2020}
\bibfield{author}{\bibinfo{person}{Phuong~T. Nguyen}, \bibinfo{person}{Juri~Di
  Rocco}, \bibinfo{person}{Davide~Di Ruscio}, {and}
  \bibinfo{person}{Massimiliano~Di Penta}.} \bibinfo{year}{2020}\natexlab{}.
\newblock \showarticletitle{CrossRec: Supporting software developers by
  recommending third-party libraries}.
\newblock \bibinfo{journal}{\emph{Journal of Systems and Software}}
  \bibinfo{volume}{161} (\bibinfo{year}{2020}), \bibinfo{pages}{110460}.
\newblock
\showISSN{0164-1212}
\urldef\tempurl%
\url{https://doi.org/10.1016/j.jss.2019.110460}
\showDOI{\tempurl}


\bibitem[\protect\citeauthoryear{Pano, Graziotin, and Abrahamsson}{Pano
  et~al\mbox{.}}{2016}]%
        {pano2016}
\bibfield{author}{\bibinfo{person}{Amantia Pano}, \bibinfo{person}{Daniel
  Graziotin}, {and} \bibinfo{person}{Pekka Abrahamsson}.}
  \bibinfo{year}{2016}\natexlab{}.
\newblock \showarticletitle{What leads developers towards the choice of a
  JavaScript framework?}
\newblock \bibinfo{journal}{\emph{CoRR}}  \bibinfo{volume}{abs/1605.04303}
  (\bibinfo{year}{2016}).
\newblock
\showeprint[arxiv]{1605.04303}
\urldef\tempurl%
\url{http://arxiv.org/abs/1605.04303}
\showURL{%
\tempurl}


\bibitem[\protect\citeauthoryear{{Piccioni}, {Furia}, and {Meyer}}{{Piccioni}
  et~al\mbox{.}}{2013}]%
        {piccioni2013}
\bibfield{author}{\bibinfo{person}{M. {Piccioni}}, \bibinfo{person}{C.~A.
  {Furia}}, {and} \bibinfo{person}{B. {Meyer}}.}
  \bibinfo{year}{2013}\natexlab{}.
\newblock \showarticletitle{An Empirical Study of API Usability}. In
  \bibinfo{booktitle}{\emph{2013 ACM / IEEE International Symposium on
  Empirical Software Engineering and Measurement}}. \bibinfo{pages}{5--14}.
\newblock
\showISSN{1949-3789}
\urldef\tempurl%
\url{https://doi.org/10.1109/ESEM.2013.14}
\showDOI{\tempurl}


\bibitem[\protect\citeauthoryear{Strauss and Corbin}{Strauss and
  Corbin}{1997}]%
        {strauss1997grounded}
\bibfield{author}{\bibinfo{person}{Anselm Strauss} {and}
  \bibinfo{person}{Juliet~M Corbin}.} \bibinfo{year}{1997}\natexlab{}.
\newblock \bibinfo{booktitle}{\emph{Grounded theory in practice}}.
\newblock \bibinfo{publisher}{Sage Publications, Inc.}
\newblock


\bibitem[\protect\citeauthoryear{Venkatesh, Thong, and Xu}{Venkatesh
  et~al\mbox{.}}{2012}]%
        {venkatesh2012}
\bibfield{author}{\bibinfo{person}{Viswanath Venkatesh}, \bibinfo{person}{James
  Y.~L. Thong}, {and} \bibinfo{person}{Xin Xu}.}
  \bibinfo{year}{2012}\natexlab{}.
\newblock \showarticletitle{Consumer Acceptance and Use of Information
  Technology: Extending the Unified Theory of Acceptance and Use of
  Technology}.
\newblock \bibinfo{journal}{\emph{MIS Quarterly}} \bibinfo{volume}{36},
  \bibinfo{number}{1} (\bibinfo{year}{2012}), \bibinfo{pages}{157--178}.
\newblock
\showISSN{02767783}
\urldef\tempurl%
\url{http://www.jstor.org/stable/41410412}
\showURL{%
\tempurl}


\bibitem[\protect\citeauthoryear{Xu, An, Thung, Khomh, and Lo}{Xu
  et~al\mbox{.}}{2020}]%
        {xu2020}
\bibfield{author}{\bibinfo{person}{Bowen Xu}, \bibinfo{person}{Le An},
  \bibinfo{person}{Ferdian Thung}, \bibinfo{person}{Foutse Khomh}, {and}
  \bibinfo{person}{David Lo}.} \bibinfo{year}{2020}\natexlab{}.
\newblock \showarticletitle{Why reinventing the wheels? An empirical study on
  library reuse and re-implementation}.
\newblock \bibinfo{journal}{\emph{Empirical Software Engineering}}
  \bibinfo{volume}{25}, \bibinfo{number}{1} (\bibinfo{year}{2020}),
  \bibinfo{pages}{755--789}.
\newblock
\showISBNx{1573-7616}
\urldef\tempurl%
\url{https://doi.org/10.1007/s10664-019-09771-0}
\showDOI{\tempurl}


\bibitem[\protect\citeauthoryear{Yano, Kula, Ishio, and Inoue}{Yano
  et~al\mbox{.}}{2015}]%
        {yano2015}
\bibfield{author}{\bibinfo{person}{Y. Yano}, \bibinfo{person}{R.~G. Kula},
  \bibinfo{person}{T. Ishio}, {and} \bibinfo{person}{K. Inoue}.}
  \bibinfo{year}{2015}\natexlab{}.
\newblock \showarticletitle{VerXCombo: An Interactive Data Visualization of
  Popular Library Version Combinations}, In \bibinfo{booktitle}{2015 IEEE 23rd
  International Conference on Program Comprehension}.
\newblock \bibinfo{journal}{\emph{2015 IEEE 23rd International Conference on
  Program Comprehension}}, \bibinfo{pages}{291--294}.
\newblock
\showISBNx{1092-8138}
\urldef\tempurl%
\url{https://doi.org/10.1109/ICPC.2015.43}
\showDOI{\tempurl}


\bibitem[\protect\citeauthoryear{Zaimi, Ampatzoglou, Triantafyllidou,
  Chatzigeorgiou, Mavridis, Chaikalis, Deligiannis, Sfetsos, and
  Stamelos}{Zaimi et~al\mbox{.}}{2015}]%
        {zaimi2015}
\bibfield{author}{\bibinfo{person}{Asimina Zaimi}, \bibinfo{person}{Apostolos
  Ampatzoglou}, \bibinfo{person}{Noni Triantafyllidou},
  \bibinfo{person}{Alexander Chatzigeorgiou}, \bibinfo{person}{Androklis
  Mavridis}, \bibinfo{person}{Theodore Chaikalis}, \bibinfo{person}{Ignatios
  Deligiannis}, \bibinfo{person}{Panagiotis Sfetsos}, {and}
  \bibinfo{person}{Ioannis Stamelos}.} \bibinfo{year}{2015}\natexlab{}.
\newblock \showarticletitle{An Empirical Study on the Reuse of Third-Party
  Libraries in Open-Source Software Development}. In
  \bibinfo{booktitle}{\emph{Proceedings of the 7th Balkan Conference on
  Informatics Conference}} (Craiova, Romania) \emph{(\bibinfo{series}{BCI
  ’15})}. \bibinfo{publisher}{Association for Computing Machinery},
  \bibinfo{address}{New York, NY, USA}, Article \bibinfo{articleno}{Article 4},
  \bibinfo{numpages}{8}~pages.
\newblock
\showISBNx{9781450333351}
\urldef\tempurl%
\url{https://doi.org/10.1145/2801081.2801087}
\showDOI{\tempurl}


\end{thebibliography}

\end{document}